\documentclass[12pt]{report}
\usepackage[letterpaper,left = 1.5in, right = 1in, top = 1in, bottom=1in, footskip = .5in]{geometry}
\usepackage{titlesec}
\usepackage{titletoc}
\usepackage{lipsum}
\usepackage{currfile}
\usepackage{physics}
\usepackage{amsmath}
\makeatletter
\def\set@curr@file#1{%
  \begingroup
    \escapechar\m@ne
    \xdef\@curr@file{\expandafter\string\csname #1\endcsname}%
  \endgroup
}
\def\quote@name#1{"\quote@@name#1\@gobble""}
\def\quote@@name#1"{#1\quote@@name}
\def\unquote@name#1{\quote@@name#1\@gobble"}
\makeatother
\usepackage{graphicx}
\usepackage{setspace}
\graphicspath{{Figures/}}
\rmfamily

\titleformat{\chapter}{\vspace{-10pt}\normalfont\bfseries}{\thechapter}{0.25in}{\centering}
\titlespacing*{\chapter}{0pt}{0in}{0pt}

\titleformat{\section}{\normalfont\bfseries}{\thesection}{0.5in}{}
\titleformat{\subsection}{\normalfont\bfseries}{\thesection}{0.5in}{}

\dottedcontents{chapter}[0pt]{\normalfont\vspace{10pt}}{10pt}{0pt}

\begin{document}

\begin{titlepage}
\begin{center}
CALIFORNIA STATE UNIVERSITY NORTHRIDGE

\vfill

Many Body Localization in 2D Systems with Quasi-Random fields in X and Y Directions

\vfill

A thesis submitted in partial fulfillment of the requirements\\ For the degree of Master of Science in Physics

\vfill

By\\
\bigskip
Nicholas A. Carrillo

\vfill
May 2019

\end{center}
\end{titlepage}
\newpage
\setcounter{page}{2}
\vspace*{\fill}
\begin{center}
Copyright by Nicholas A. Carrillo 2019
\addcontentsline{toc}{chapter}{Copyright}
\end{center}

\newpage

\addcontentsline{toc}{chapter}{Signature Page}
\setcounter{page}{3}
The thesis of Nicholas A. Carrillo approved:
\vfill

\rule{8cm}{0.4pt}\hfill\rule{3cm}{0.4pt}\\
Dr. Yohannes Shiferaw\hfill Date
\vfill

\rule{8cm}{0.4pt}\hfill\rule{3cm}{0.4pt}\\
Dr. Say-Peng Lim\hfill Date
\vfill

\rule{8cm}{0.4pt}\hfill\rule{3cm}{0.4pt}\\
Dr. Donna Sheng, Chair\hfill Date 
\vfill

\begin{center}
California State University, Northridge
\end{center}

\newpage

\chapter*{\normalfont Acknowledgments}
\addcontentsline{toc}{chapter}{Acknowledgments}
\bigskip
The author would like to thank both Dr. Donna Sheng, and Dr. Say-Peng Lim for all the help and guidance they have provided over the years in conducting research. That guidance was continued in the formation of this manuscript, and the research involved with this thesis. The author would also like to acknowledge that it was Dr. Donna Sheng, and Dr. Say-Peng Lim who determined the underlying cause of the level clustering seen in the results, for which the author is incredibly grateful. 

\tableofcontents

\chapter*{\normalfont List of Figures}
\addcontentsline{toc}{chapter}{List of Figures}
\bigskip
Figure 1.1 (pg. 6): Here we show how \textbf{r} changes as a function of increasing disorder amplitude for the disordered-Heisenberg model (eq. 1.5). For each disorder amplitude value, $h$ we iterate over 1000 $h_i = [-h,h]$ for the {\normalfont 4x3}, and 200 for the {\normalfont 5x3} system. We see \textbf{r} decrease to its Poisson value in the limit $\frac{J}{h}<<1$, indicating that the system's states are becoming more localized.\\

Figure 3.1 (pg. 16): Here we show the entanglement entropy per site for different system sizes, for the quasi-Heisenberg model. Here we see a drop in EE as the system becomes more localized with increasing h, indicating an MBL phase.\\

Figure 3.2 (pg. 17): Plot of EE distributions with peaks forming around $S=0$, and $S=ln(2)$ when a system enters the MBL phase, as disorder amplitude is increased. The different plots are for different system sizes, both showing a peak forming around $ln(2)$ for increasing disorder amplitude\\

Figure 3.3 (pg. 18): Plots comparing the adjacent gap ratios for the {\normalfont 4x3, 5x3, 6x3} lattices as a function of disorder amplitude; the horizontal line in both is the Poisson limit. The top plot is for quasirandom disorder, and the bottom plot is for uniform disorder. Note: \textbf{r} for the {\normalfont 6x3} system was only calculated for 9 disorder values, and only for quasirandom disorder.\\

Figure 3.4 (pg. 18): P(r) for different system sizes, with uniform disorder, showing the expected transition from a GOE, to Poisson distribution as we increase the disorder amplitude. We also fit these distributions with eq. 3.2, where $\omega = 1(0)$ gives the GOE(Poisson) distribution.\\

Figure 3.5 (pg. 19): Plot of the fitted Brody parameters as a function of disorder amplitude. The different curves are for different system sizes. The plot on the left is for quasirandom disorder, where the one on the right is for uniform disorder. We see $\omega$ reflect the expected behavior for \textbf{r} in the right plot, and it tends towards its Poisson value ($\omega = 0$).\\

Figure 3.6 (pg. 20): P(r) for different nx3 lattice sizes, with each plot showing P(r)for different disorder amplitudes with corresponding Brody parameter ($\omega$). The right column shows P(r) for the extreme disorder value h=30.0. For the left column, we show a transition from a GOE , to Poisson, to a level clustering distribution. We notice in the extreme values for h, the distribution sharply peaks for small values of r, indicating most levels are clustered for higher disorder values. We also note that as total system size increases, the clustering at h=30.0 is less extreme.\\

Figure 3.7 (pg. 21): P(r) for 6x2, and 4x4 lattice, with each plot showing P(r) for different disorder amplitudes with corresponding Brody parameter ($\omega$). The right column shows P(r) for the extreme disorder value h=30.0. We still see a transition from a GOE, to Poisson, to level clustering distribution as h increases, and again we the clustering is less extreme for h=30.0, for larger system sizes. Despite a different geometry to that of the nx3 lattice, the behavior is similar, indicating an independence of geometry.

\chapter*{\normalfont Abstract}
\addcontentsline{toc}{chapter}{Abstract}

\bigskip
\begin{center}
Many Body Localization in 2D Systems with Quasi-Random fields in X and Y Directions\\
\bigskip

By
\bigskip

Nicholas A. Carrillo\\
\bigskip

Master of Science in Physics\\
\bigskip
\end{center}

\begin{flushleft}
\linespread{1.6}\selectfont Many body localization (MBL) is a phenomena that allows for the preservation of quantum information for long times. We study a variation of the disordered-Heisenberg model, which is known to exhibit an MBL phase$^{[5][6]}$, known as the quasi-Heisenberg model. Our model is a variation of the quasi-Heisenberg model with fields in both x and y directions. We found that while our model shares some characteristics for MBL, as seen by other quasi-Heisenberg models, the adjacent gap ratio for our system falls well below the expected Poisson limit when it transitions to an MBL phase. A similar model to ours has been experimentally realized in [11], and so we are motivated further to study MBL characteristics. To determine the characteristics of the system when transitioning from the ergodic to MBL phase, we calculate the quantities: entanglment entropy, adjacent gap ratio, and the probability distributions for both. We calculate the aforementioned quantities through the eigenvalues and eigenvectors of the system's Hamiltonian, which are obtained through exact diagonalization. We find the entanglement entropy of the system behaves as expected for MBL, with the spectral average of the entanglement entropy dropping close to zero as it enters the MBL phase. However, the spectral average for the adjacent gap ratio falls below the Poisson limit, 0.386, which corresponds to uncorrelated energy levels. This drop in the adjacent gap ratio can be explained through some level of attraction between energy levels, and is ultimately a consequence of the separability of the quasirandom field. The probability distribution for the adjacent gap ratio is fitted from a derived distribution which was found using the Brody function for energy level spacing. The Brody function accounts for energy level attraction when the Brody parameter, $\omega$, becomes negative. We find by just varying this single parameter, we can accurately fit the distribution to our data, indicating level clustering when $\omega < 0$ is used to fit the distribution for the adjacent gap ratio.
\end{flushleft}

\chapter{Introduction}

\pagenumbering{arabic}

\bigskip

The following discussion summarizes what is extensively discussed in sources [1],[2], and [4]. Statistical mechanics tells us that if we have a subsystem coupled to a reservoir, the state of the system after a long time can be described by the micro-canonical, canonical, and grand-canonical ensembles, and are said to \textit{thermalize} with the reservoir. We can consider whether a closed system can act as its own reservoir; that is, the system can act as a reservoir for distinct subsystems within itself. We apply this idea to quantum statistical mechanics, and consider a closed quantum system. It should be noted that if a measuring apparatus is also a quantum system, then it too could be considered part of the larger closed quantum system. With a system acting as its own reservoir, we expect a subsystem which it shares a few degrees of freedom with to couple to the rest of the system. In coupling to the rest of the system we expect after long times for the subsystem to thermalize with the entire system. However, there is class of closed quantum systems which do not thermalize their subsystems; they are known as many-body Anderson localize systems, or MBL systems.\\
\bigskip

Since MBL systems do not thermalize, initial, localized details of a system's initial state do not average out over long periods of time, i.e. their states do not decohere. The preservation of information about the initial state of the system makes MBL systems a candidate for quantum memory applications$^{[1]}$. Interestingly, it has been shown that MBL occurs in highly excited states for interacting many-body quantum systems with strong coupling to an external random field, indicating that some MBL systems are robust under random fluctuations to the system$^{[1]}$.\\
\bigskip

\section{Many-Body Quantum Mechanics}
\bigskip

We are interested in Hamiltonians that are localized and time-independent, and so we focus on those with short-range, nearest-neighbor interactions, and on-site coupling to some disorder field, and in our case a spatially varied magnetic field. We also focus on highly excited states, away from the ground state of the system, and observe the dynamics of such states. The eigenstates of the system are defined by:

\begin{center}
$H\ket{n} = E_n\ket{n}$ 
\end{center}

where $\ket{n}$ can be expanded in some basis ${\ket{\phi}}$ as:

\begin{center}
$$\ket{n} = \sum_{\ket{\phi}} \bra{\phi}\ket{n}\ket{\phi}$$
\end{center}

We can analyze these states using the formalism of probability (density) operators. The dynamics of a state can be studied, using the Schrodinger representation where the operator evolves in time according to:

\begin{center}
$\rho (t) = e^{-\textit{i}Ht/\hbar}\rho (0)e^{\textit{i}Ht/\hbar}\hspace{2cm}\textit{i}\hbar \frac{d\rho}{dt} = [H,\rho]\hspace{2cm}Tr(\rho) = 1 \hspace{0.5cm} (1.1)$
\end{center}

other operators, \textbf{O}, are time independent, and their expectation values as a function of time can be given as, $<\textbf{O}(t)> = Tr(\textbf{O}\rho (t))$.\\
\bigskip

The basis states we work with represent a system of spin-$\frac{1}{2}$ particles on a lattice, with each spin having a Hilbert space dimension of $(2S + 1)$, or 2 for spin-$\frac{1}{2}$ particles. Obviously the Hilbert space dimension for the full system is significantly larger than that for an individual spin. A convenient, and conventional, way to construct a full state of the many-body system, $\ket{S_z}$, is by taking consecutive outer products of each $S_z$ state for each spin, $\ket{\phi_i^z}$, as shown below:

\begin{center}
$ \ket{S_z} = \ket{\phi_{1}^z}\otimes\ket{\phi_{2}^z}\otimes...\otimes\ket{\phi_{N-1}^z}\otimes\ket{\phi_N^z} = \ket{\phi_1^z\phi_2^z...\phi_{N-1}^z\phi_N^z}$
\end{center}

The $2^N$ basis states for the system is simply all the outer product combinations of eigenstates for the single-particle $S_i^z$ operators, i.e. the eigenstates of the total $S_z$ operator. Operators expanded to the full Hilbert space of the system can be formulated in much the same way, by taking consetutive outer products of single-spin operators. In fact, each spin has four linearly independent operators that can act on it: the identity operator, and the three Pauli spin matrices$^{[1]}$. As an example, the $S_z$ operator for the ith spin expanded to the full Hilbert space of the system can be constructed as:

\begin{center}
$S_i^z = \underbrace{I\otimes I\otimes ... \otimes I\otimes}_{i-1} \sigma_z\underbrace{\otimes I\otimes ... \otimes I\otimes I}_{N-i}$ 
\end{center}

where N is the total number of spins, and the $i$th spin contributes its Pauli spin matrix, $\sigma_z$, and all the other spins contribute their identity operators, $I$. The Hamiltonian we are concerned with is a sum of local operators, i.e. spins only interact with nearest-neighbors, and or on-site couplings, such as the $S_i^z$ operator, or $\textbf{S}_i\cdot\textbf{S}_j$, where i and j are nearest-neighbors.\\

We will use the eigenstates of total $S_z$ as the basis states for which we construct our Hamiltonian. In this way, the full many-body eigenstates of the Hamiltonian are a linear combination of total $S_z$ eigenstates, and it is these full many-body eigenstates that are used to construct $\rho$.
\bigskip

\section{Quantum Thermalization}
\bigskip

When subsystems thermalize to their reservoirs, they take on the extensive quantities of their reservoirs, and any information about the initial state of the subsystem is lost. This is an \textit{apparent} paradox since if we time evolve some initial state of the subsystem through a unitary transformation, no information about the initial state is lost. The paradox is rectified if we consider the information to be initially localized, as in a localized state, but then it diffuses through the entire system over a long time$^{[1][2]}$. Consider a situation where information for some localized state does not diffuse. Then if we wish to retrieve that information we must make a measurement of some local operator. However, a lack of information spreading means the system does not thermalize. But if we wish to access the information of the initial, localized state in a thermalized system, it would require the measurement of some global operator, and would therefore be irretrievable.\\

The spreading of localized states allows closed quantum systems to act as reservoirs for their subsystems, and is a main focus of equilibrium quantum statistical mechanics. To understand the thermalization of a subsystem (A), we consider a system with energy being the only conserved extensive quantity, with that energy pertaining to some temperature. If the system is to act as its own reservoir, then the system's Hamiltonian (\textbf{H}) must connect all degrees of the freedom within A, to the rest of the system, B.
\bigskip

Given that the state of the system is represented by $\rho (t)$, we can extract $\rho_A(t)$, known as the reduced density matrix, from $\rho(t)$ by taking a partial trace$^{[1][2]}$:

\begin{center}
$\rho_A = Tr_B(\rho (t))\hspace{.5cm}(1.2)$
\end{center}

We consider that full many-body states can be Schmidt decomposed into a tensor product state: $\ket{\Phi} = \ket{A}\otimes\ket{B}$, with a trace over B giving us the reduced density matrix for system A. The $\rho (t)$ that we extract $\rho_A(t)$ from reaches an equilibrium distribution of: 

\begin{center}
$\rho^{eq}(T) = Z^{-1}e^{-H/k_bT}$
\end{center}

and so $\rho_A^{eq} = Tr_B{\rho^{eq}(T)}$ where system A thermalizes (in the long time, large system limit) when $\rho_A(t) = \rho_A^{eq}(T)$. It is then reasonable to assume that if an initial state does thermalize for a given temperature (energy), then all states at that energy thermalize. In the context of thermalization, we are interested in states that pertain to higher energy densities, far away from the ground state.$^{[1]}$.
\bigskip

\section{The Eigenstate Thermalization Hypothesis(ETH)}
\bigskip

We can look at thermalization in the context of the system's many body eigenstates, at which point the state at long times is trivial: $\rho (0) = \rho (t)$. If all initial states thermalize at a corresponding temperature (T), then every many-body eigenstate corresponding to that energy, temperature, will also thermalize. This implies that \textit{all many-body eigenstates are thermal}, this is known as the Eigenstate Thermalization Hypothesis (ETH). To clarify, since the eigenstates are defined by the solution to:

\begin{center}
$\textbf{H}\ket{n} = E_n\ket{n}$
\end{center}

and $\rho (0) = \rho (t)$, then the eigenenergy corresponds to the thermal energy associated with the state at equilibrium, and is denoted as $T_{eq}^n$. If the state of the system can be represented by the eigenstate $\ket{n}$ as:

\begin{center}
$\rho_n = \ket{n}\bra{n}$
\end{center}

then the reduced density matrix of A is given by:

\begin{center}
$\rho_A^n = Tr_B(\rho_n)\hspace{0.5cm}(1.3)$
\end{center}

In accordance with the ETH, the system thermalizes as $\rho_n = \rho (T_{eq}^n)$ and $\rho_A^n = \rho_A(T_{eq}^n)$. One consequence is that the entanglement entropy between A and B is given by:

\begin{center}
$S_{AB} = -k_BTr_B(\rho_A^nln(\rho_A^n))\hspace{0.5cm} (1.4)$
\end{center}

and is equal to the entropy of system A after it thermalizes. This entropy between systems A and B is proportional to the volume of subsystem A, and so the entanglement entropy for, thermal eigenstates, between the two systems obeys volume-law scaling$^{[1]}$. The ETH is not true, however, for systems that are many-body Anderson localized.
\bigskip

\section{Localization}
\bigskip

Lets look at the disordered-Heisenberg model for a system with nearest neighbor interaction, and an on-site coupling to a random magnetic field:

\begin{center}
$$H = J\sum_{<ij>} \textbf{S}_i \cdot \textbf{S} _j + \sum_i h_iS_i^z \hspace{0.5cm} (1.5) $$
\end{center}

From the model, if J is set to zero, then the many-body eigenstates are simply the product states of spins, $\ket{\sigma_i^z}$, and the entire system is localized since there are no interactions to diffuse any information about the initial state. We can consider weak interactions $(h>>J)$, and perturbatively construct the many-body eigenstates, and for sufficiently strong disorder, the system remains more or less localized, and does not obey the ETH. The perturbative argument is restrained to relatively weaker interactions, but for relatively stronger ones we still see the system obey the ETH for weak disorder. However for stronger interactions, well outside of perturbation theory, and stronger disorder the system fails to obey the ETH. This failure to obey the ETH has been shown numerically$^{[6][12]}$, with a transition to the MBL phase around $h = 3.5$ for the 1-D system of (1.5). This behavior is seen even for states with higher energy density ($\epsilon = 0.5$), well above the ground state, and it is these highly excited states that we are interested in. The energy density ($\epsilon$) is given by:

\begin{center}
$$\epsilon = \frac{E_{max}-E}{E_{max}-E_{min}}$$
\end{center}

where $E_{max}$($E_{min})$ is the maximum(minimum) energy eigenvalue for a Hamiltonian, and $E$ is some value between them. A variation of the above model is known as the quasi-Heisenberg model, and is given by:

\begin{center}
$$H = J\sum_{<ij>} \textbf{S}_i \cdot \textbf{S} _j + h\sum_i S_i^zcos(2\pi cn_i + \phi)\hspace{0.5cm}(1.6)$$
\end{center}

where $c$ is an irrational number, and $n_i$ is the site index in the 1-D case. This model has also been shown to exhibit MBL for states with higher energy density$^{[5]}$, and a 2-D version is the focus of this study. 
\bigskip

\section{Ergodic to MBL Phase Transition}
\bigskip

We have discussed that certain models exhibit a transition to an MBL phse, but how to determine whether a system is in the MBL phase requires the calculation of the various quantities: entanglement entropy,and the adjacent gap ratio and its probability distribution. Based on the quasi-Heisenberg model, for weak disorder amplitudes we find that the many-body eigenstates obey the ETH, however after a certain critical disorder amplitude $(h_c)$, the system' eigenstatres no longer obey the ETH as the system enters the MBL phase.
\bigskip

\subsection*{Adjacent Gap Ratio (AGR)}
\bigskip

The AGR was a quantity introduced by Oganesyan and Huse$^{[3]}$ to probe the spectral statistics of the system's eigenstates. As the system enters the MBL phase, we expect nearby eigenstates to become uncorrelated, non-interacting, so as to preserve the localization of states. We can look at the difference in neighboring eigen-energies that have been sorted by increasing energy as a way to gauge the spectral statistics of the eigen-energies, namely:

\begin{center}
$\delta_n = E_{n+1} - E{n}\hspace{2cm}r_n = min(\delta_{n},\delta_{n-1})/max(\delta_n,\delta_{n-1})$
\end{center}

with $r_n$ being the adjacent gap ratio. This quantity has the benefit of only relying on differences of energies, and not on absolute values. As a quick note, we drop the $n$ subscript and use $r$ with the understanding that it is still a discrete quantity, and the spectral average, $[<r>] = \textbf{r}$, is what is ultimately calculated, as shown in Fig. 1.1. In our case $<r>$ is the average value for a spectrum of eigenvalues, and $[<r>]$ is average over different spectra given by varying $\phi$ while keeping $h$ constant in eq. 1.6, for example.\\
\bigskip

\begin{figure}[!htb]
\begin{center}
\includegraphics[scale=0.4]{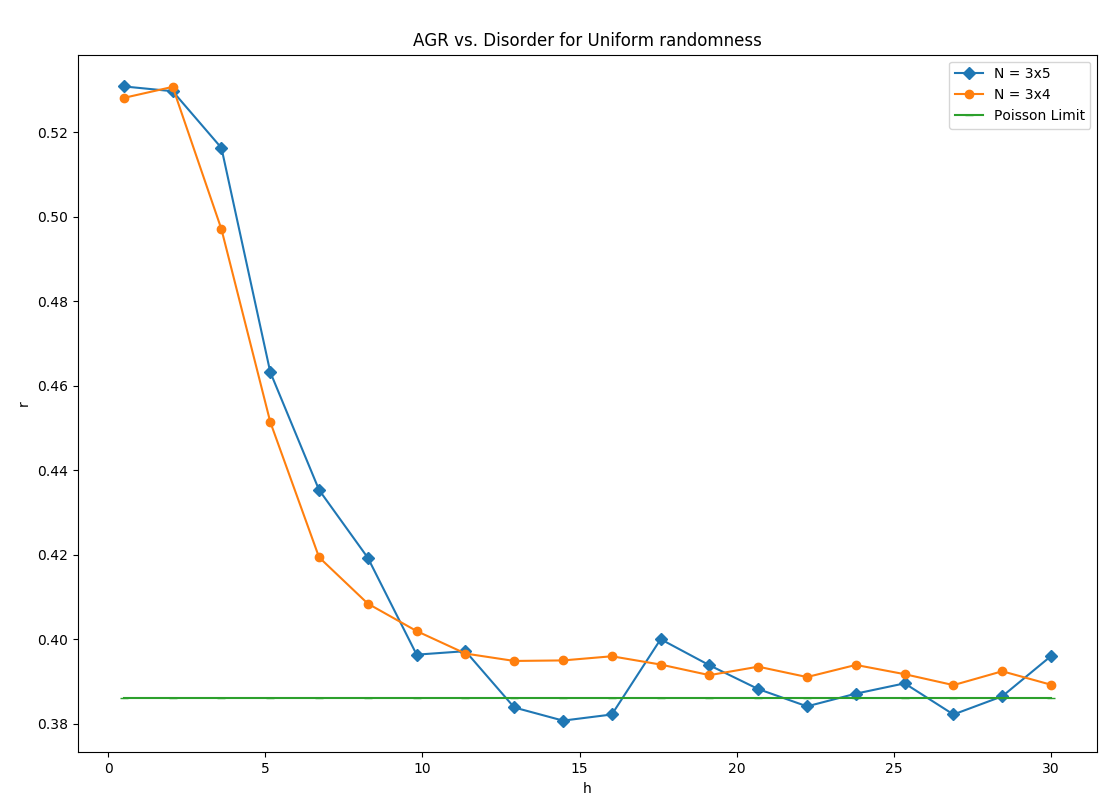}
\end{center}
\caption{\textit{Here we show how \textbf{r} changes as a function of increasing disorder amplitude for the disordered-Heisenberg model (eq. 1.5). For each disorder amplitude value, $h$ we iterate over 1000 $h_i = [-h,h]$ for the {\normalfont 4x3}, and 200 for the {\normalfont 5x3} system. We see \textbf{r} decrease to its Poisson value in the limit $\frac{J}{h}<<1$, indicating that the system's states are becoming more localized.}}
\end{figure}

For the MBL phase we expect the eigenstates to be uncorrelated, and the probability distribution of the $r$ to be Poisson distributed in the limit as the system size goes to infinity$^{[3]}$. In this limit, the distribution is given as, $P(r) = 2/(1+r^2)$, and the average value of $r$ is given as, $<r> \cong 2ln(2) - 1 \cong 0.386$. In the ergodic phase, $P(r)$ can be numerically determined from a Gaussian Orthogonal Ensemble (GOE) of random matrices, with $<r> \cong 0.5295 \pm 0.0006^{[3]}$. In plotting \textbf{r} as a function of disorder amplitude ($h$), and $P(r)$ we can observe a transition from the ergodic phase, to the MBL phase. We expect \textbf{r} to continuously change from its ergodic value ($\textbf{r} = 0.5295$) to its Poisson value ($\textbf{r} = 0.386$), and for $P(r)$ to go from a GOE distribution, to a Poisson distribution.

\subsection*{Entanglement Entropy}
\bigskip

The entanglement entropy (EE) for two subsystems represents a spreading of information between them, and as a system becomes localized the spread of information between subsystems diminishes. If we look at our model (1.6) in the limit as $\frac{J}{h} \rightarrow 0$ the system becomes fully localized as there are no longer interactions between spins. We therefore expect a decrease in entanglement entropy as we increase the value for $h$, which is characteristic for the MBL phase. The EE is found by taking the partial trace over the density matrix, $\rho_n = \ket{n}\bra{n}$ in accordance with (1.3), and since $\rho_A$ is diagonalizable, we can rewrite (1.4) as:

\begin{center}
$$S_{AB} = -\sum_{\lambda_A} \lambda_A ln(\lambda_A)$$
\end{center}

with a summation over the eigenvalues, $\lambda_A$, of $\rho_A$. Therefore in order to calculate the EE of the system we simply need to construct $\rho_A$, and find its eigenvalues. Performing a partial trace over system B's states is somewhat unclear, and is best shown through an example between two spins. Possible states of the system can be represented by a superposition of the basis states:

\begin{center}
$\ket{\uparrow\uparrow}, \ket{\uparrow\downarrow},\ket{\downarrow\uparrow},\ket{\downarrow\downarrow}$ 
\end{center}

Where the first spin is system A, and the second spin is system B, with possible combinations:

\begin{center}
$\ket{\uparrow\uparrow},\ket{\downarrow\downarrow}$\\
$\ket{\uparrow\uparrow} \pm  \ket{\downarrow\downarrow}$\\
$\ket{\uparrow\downarrow} \pm \ket{\downarrow\uparrow}$
\end{center}

Let's calculate the EE for $\rho = (\ket{\uparrow\uparrow} \pm  \ket{\downarrow\downarrow})\cdot(\bra{\uparrow\uparrow} \pm  \bra{\downarrow\downarrow})$:

\begin{center}

$$\rho_A = \sum_{\ket{B}} \frac{1}{2}\bra{B}(\ket{\uparrow\uparrow} \pm  \ket{\downarrow\downarrow})\cdot(\bra{\uparrow\uparrow} \pm  \bra{\downarrow\downarrow})\ket{B}\hspace{0.5cm};\hspace{0.5cm} \ket{B} = \ket{\uparrow}_B,\ket{\downarrow}_B$$
\medskip

$\rho_A = \frac{1}{2}(\ket{\uparrow}\bra{\uparrow} + \ket{\downarrow}\bra{\downarrow})$ with eigenvalues $ \lambda_A = \frac{1}{2}, \frac{1}{2}$
\end{center}

Note we get the same $\rho_A$ for the state $\ket{\uparrow\downarrow} \pm \ket{\downarrow\uparrow}$. Plugging this into our equation for EE, we have:

\begin{center}
$S_{AB} = \frac{1}{2}ln(2) + \frac{1}{2}ln(2) = ln(2)$
\end{center}

This $ln(2)$ value represents the maximum entanglement between two spins. We now repeat the above calculation for $\rho = \ket{\uparrow\uparrow}\bra{\uparrow\uparrow}$:

\begin{center}
$$\rho_A = \sum_{\ket{B}}\bra{B}(\ket{\uparrow\uparrow}\bra{\uparrow\uparrow})\ket{B}$$\\
\medskip

$\rho_A = \ket{\uparrow}\bra{\uparrow}$ with eigenvalues $\lambda_A = 1,0$\\
\medskip

$S_{AB} = ln(1) + 0ln(0) = 0$ 
\end{center}

Here we see no EE for when system is in the state $\ket{\uparrow\uparrow}(\ket{\downarrow\downarrow})$. For the MBL phase there is very limited, none for the fully many-body localized (FMBL) regime,  spreading of information for localized states. So unlike the thermal phase where spins may indirectly interact with one another, e.g. one spin interacts with a third through an intermediary second spin. In the localized phase there are only direct, non-zero probability interactions between spins, where the interaction strength falls off exponentially with distance. Consequences of this localization is that the entanglement entropy grows logarithmically in time, and follows area-law scaling, as opposed to volume-law scaling$^{[1]}$.

\newpage

\chapter{Methods}

\bigskip

Exploring the MBL phase with exact diagonalization in 2D systems has proven to be more difficult than the 1D case since the Hamiltonian matrices are more dense as there are now two interactions per site. This leads to long computational times for solving the eigenvalues-eigenvectors for the Hamiltonian, and so require that we are efficient with our calculations. In Chapter 1 we discussed how the Hamiltonian can be written as a sum of Spin operators expanded to the full Hilbert space. There are two issues with this method: One is simply using matrix multiplication to generate the Hamiltonian is computationally inefficient and takes a long time. The second issue is that performing exact diagonalization on the full Hamiltonian is also incredibly time consuming, and often we only want to analyze states with some total $S_z$ value, e.g. $S_z = 0$. We tackle both of these issues by determining how the Hamiltonian acts on the set of basis states with some total $S_z = j$, and generate a block of the Hamiltonian where all of its eigenstates have total $Sz=j$ through element wise construction. This manner of construction allows us to efficiently construct smaller, ultimately sparse, block Hamiltonians, where exact diagonalization takes a reasonable amount of time.\\
\bigskip

We know that for systems with nearest neighbor interaction, and on-site coupling to a random field, we observe an MBL transition as the strength of the random field is increased$^{[6]}$. The transition is characterized by the entanglement entropy (EE), the values for \textbf{r}, and the change in the distribution for P(r). As the system transitions into the MBL phase, we see a drop in the average EE, and \textbf{r} values. The probability distribution for the adjacent gap ratio(AGR) transitions as well, from a GOE distribution to a Poisson distribution, with the transition to a Poisson distribution for P(r) being indicative of uncorrelated energy levels.\\
\bigskip

\section{The Model}
\bigskip

The disordered Heisenberg model is characterized with coupling to a uniformly random field, thus having uniform disorder, has been shown to exhibit an MBL phase at higher disorder values for both one and two dimensional systems$^{[5][6]}$. In this study we look at a variation of the quasi-Heisenberg model for an n-leg ladder system of $1/2$-spins; where only one spin is allowed at each site and is either spin-up or spin-down. Again, the Heisenberg model for nearest-neighbor interactions is given by:
\begin{center}
$$H = J\sum_{<ij>} \textbf{S}_i\cdot\textbf{S}_j + \sum_i h_iS_i^z \hspace{0.5cm}(2.1)$$ \\
\end{center}
where $<ij>$ denotes a sum over nearest neighbors, and $h_i$ is the random disorder amplitude.\\

The quasi-Heisenberg model refers to an cosine term attached to the on-site coupling with an irrational wavelength, this makes the field incommensurate and introduces a quasi-randomness to the model which is given by:
\begin{center}
$$H = J\sum_{<ij>} \textbf{S}_i\cdot\textbf{S}_j + h\sum_i S_i^zcos(2\pi cn_i + \phi)\hspace{0.5cm}(2.2)$$ \\
\end{center}

where $J$, and the disorder amplitude, $h$, are constant, $n_i$ refers to the site index in the 1D case, $c$ is an irrational number, e.g. $\sqrt{2}$, and $\phi$ is some random phase between $[0,\pi]$. For the 1D case $n_i$ is simply the site-index. The model we explore is a variation of the quasi-Heisenberg model, and is given by:

\begin{center}
$$H = J\sum_{<ij>} \textbf{S}_i\cdot\textbf{S}_j + h\sum_i S_i^z[cos(2\pi cn_i^x + \phi ') + cos(2\pi cn_i^y + \phi)]\hspace{0.5cm}(2.3)$$
\end{center}

In this model we have two separate fields one along the $x$-direction, and another along the $y$-direction. Where $n_i^x$($n_i^y$) refers to the row(column) index for the $i$th site; both $\phi '$ and $\phi$ are random phases between $[0,\pi]$. In other 2D models, such as in [5], there is a single cosine term with $n_i$ typically referring to either the row or column index of the site, and $\phi$ is constant for either an entire row, or column.\\

\section{Generating the Hamiltonian Matrix}
\bigskip

The Hamiltonian can be written as a sum of interaction, and spin operators expanded to the full Hilbert space with the interaction operators simply being two spin operators multiplied together. The spin operators can be constructed, as an example, via:

\begin{center}

$$S_i^z = \underbrace{I\otimes I\otimes ... \otimes I\otimes}_{i-1} \sigma_z\underbrace{\otimes I\otimes ... \otimes I\otimes I}_{N-i}\hspace{0.5cm}(2.4)$$ 

\end{center}

Where $\sigma_z$ is the Pauli spin matrix, $i$ is the particle index, and $N$ is the total number of particles. Tensor products, while straightforward, are computationally time consuming since every element of one matrix is multiplied to \textit{every other} element of the other matrix. Given that the full Hilbert space dimension is given as $2^N$ the tensor product formulation quickly becomes unwieldy.\\
\bigskip

We learned from Sakurai$^{[14]}$ that a matrix, namely the Hamiltonian, can be constructed element wise in a given basis $\ket{i}$ via:

\begin{center}
$$ H' = \sum_{i,i'}\ket{i}\bra{i}H\ket{i'}\bra{i'} \hspace{0.5cm}(2.5)$$ 
\end{center}

If we understand how $H$ acts on our basis states then we can find each element $H_{ij}$. For this system we find that the total $S_z$ operator commutes with the Hamiltonian, and so we can use the $S_z$ eigenstates to construct our Hamiltonian. These eigenstates are consecutive tensor products of each site's spin state $\ket{\phi_i^z}$, given by:
\bigskip

\begin{center}
$\ket{S_z} = \ket{\phi_{1}^z}\otimes\ket{\phi_{2}^z}\otimes...\otimes\ket{\phi_{N-1}^z}\otimes\ket{\phi_N^z} = \ket{\phi_1^z\phi_2^z...\phi_{N-1}^z\phi_N^z} \hspace{0.5cm} (2.6)$\\
\end{center}
\bigskip

For a single $i$th particle, its $S_i^x$ and $S_i^y$ operators can be expressed in terms of raising and lowering operators, from Sakurai we have: $S_i^+ = S_i^x + \textit{i}S_i^y$ and $S_i^- = S_i^x - \textit{i}S_i^y$. From here:

\begin{center}
$S_i^+ + S_i^- = 2S_i^x$ and $S_i^+ - S_i^- = 2\textit{i}S_i^y\hspace{0.5cm} (2.7a)$\\
\medskip

$S_i^x = \frac{1}{2}(S_i^+ + S_i^-)$ and $S_i^y = \frac{1}{2\textit{i}}(S_i^+ - S_i^-)\hspace{0.5cm} (2.7b)$
\end{center}

Ignoring the disorder term for now, we look at the interaction term $S_i \cdot S_j$:
\begin{center}
$S_i \cdot S_j = S_i^zS_j^z + S_i^xS_j^x + S_i^yS_j^y \hspace{0.5cm} (2.8a)$\\
\medskip
$S_i \cdot S_j  = S_i^zS_j^z + \frac{1}{2}(S_i^+S_j^- + S_i^-S_j^+)\hspace{0.5cm} (2.8b)$
\end{center}

This is the Jordan-Wigner transformation and we can use it to easily see how our Hamiltonian acts on our product states. Since operators like $S_i^z$ only acts on the $i$th particle, and all particles have the same same total spin, i.e. $j = \frac{1}{2}$, we can find how the operators act on our states. From Sakurai we have:

\begin{center}
$S_i^+\ket{m_1...m_i...m_N} = \hbar \sqrt{(\frac{1}{2}-m_i)(\frac{1}{2}+m_i+1}\ket{m_1...m_i+1...m_N}\hspace{0.5cm} (2.9a)$\\
\medskip

$S_i^-\ket{m_1...m_i...m_N} = \hbar \sqrt{(\frac{1}{2}-m_i)(\frac{1}{2}-m_i+1}\ket{m_1...m_i-1...m_N} \hspace{0.5cm} (2.9b)$\\
\medskip

$S_i^z\ket{m_1...m_i...m_N} = \hbar m_i\ket{m_1...m_{i}...m_N}\hspace{0.5cm}  (2.9c)$
\end{center}

Using the above relations in combination with the Jordan-Wigner transformation we have:

\begin{center}
$\frac{1}{2}S_i^+S_j^-\ket{m_1...m_i...m_N} = \frac{\hbar^2}{2}\ket{m_1...m_i+1...m_j-1...m_N} \hspace{0.5cm} (2.10a)$\\ 
\medskip

$\frac{1}{2}S_i^-S_j^+\ket{m_1...m_i...m_N} = \frac{\hbar^2}{2}\ket{m_1...m_i-1...m_j+1...m_N} \hspace{0.5cm} (2.10b)$\\ 
\medskip

$S_i^zS_j^z\ket{m_1...m_i...m_j...m_N} = \hbar^2 m_im_j\ket{m_1...m_{i}...m_j...m_N}\hspace{0.5cm} (2.10c)$
\end{center}

where $m_i$ is the spin of the $i$th particle. Upon immediate inspection we see that if both sites have the same spin direction then the coefficient for $S_i^+S_j^-$($S_i^-S_j^+$) will be zero, and the combination of raising and lowering terms simply pertains to a spin-flip between sites $i$ and $j$.\\
\bigskip

Here is a good time to introduce binary representation for our states. Given that each site only has two possible states, we can label $m = \frac{1}{2}$ as 1, and $m = -\frac{1}{2}$ as 0, e.g. $\ket{1010}$ is the state of a 4-spin system with alternating spin directions. We are now in a position to generate our Hamiltonian through element wise construction.

\section{2D Square Lattice}
\bigskip

For the 2D system, adjacent spins do not necessarily correspond to adjacent indices. As an example, we can generate the Hamiltonian for a 4-spin system with open boundary conditions in either direction. We label the spins as shown below:

\begin{center}
1$\bullet$\hspace{1cm}2$\bullet$\\

\vspace{1cm}
3$\bullet$\hspace{1cm}4$\bullet$
\end{center} 

Again, we are only considering nearest neighbor interactions so first we generate the states, and make note of which site indices are nearest neighbors. We see that our nearest neighbor pairs are:

\begin{center}
(1,2),(1,3),(2,4),(3,4)
\end{center}

We must be careful not to double count our interactions in our open system, e.g. (1,3) and (3,1) is the same interaction. To generate our states we can choose a decreasing binary representation to represent our $S_z$ eigenstates:

\begin{center}
$\ket{1111},\ket{1110},\ket{1101},\ket{1011},...,\ket{0010},\ket{0001},\ket{0000}$ for a total of 16 states.
\end{center}

We then make note of which sates interact via the Hamiltonian, e.g. $\ket{0000}$ doesn't undergo any spin flipping, whereas other states are connected via spin flipping:

\begin{flushleft}
	\hspace{5cm}$\ket{0000}\rightarrow\ket{0000}$\\
	\hspace{5cm}$\ket{0001}\rightarrow\ket{0001},\ket{0100},\ket{0010}$\\
	\hspace{5cm}\vdots
\end{flushleft}

We notice that the state $\ket{0001}$ is connected to two other states via our Hamiltonian. Since we have an ordered basis we can label each state with an index, in this case the above states have index labels:

\begin{flushleft}
	\hspace{5cm}$\ket{0000} = 16$\\
	\hspace{5cm}$\ket{0001} = 15$\\
	\hspace{5cm}$\ket{0100} = 12$\\
	\hspace{5cm}$\ket{0010} = 14$\\
\end{flushleft}

We note that all connected states correspond to non-zero matrix elements with coordinates corresponding to the connected state indices. Based on the above example we see that:

\begin{center}
H(16,16),H(15,15),H(15,14),H(15,12)
\end{center}

Are all non-zero matrix elements, with coefficients determined from equations 2.10.

From here we simply iterate over each sate, and then make note of which states are connected to each other. In this way, we generate all non-zero matrix elements with their respective matrix coordinates. And so we have shown that we can generate the Hamiltonian via element-wise construction, thus avoiding tensor products and matrix multiplication altogether. In fact based on the above scheme, we only ever deal with non-zero matrix elements which allows us to construct a sparse matrix based on the coefficients and their coordinates, the exact diagonalization of which is much more efficient.
\bigskip

\section{Adding Disorder}
\bigskip

We now wish to add the disorder term to our interaction term utilizing the framework of our binary basis representation. If we only look at the disorder term:

\begin{center}
$$H_{disorder} = h\sum_i S_i^z[cos(2\pi cn_i^x + \phi ') + cos(2\pi cn_i^y + \phi)]\hspace{0.5cm} (2.11)$$
\end{center}

We immediately notice that it is an on-site interaction, i.e. no connected states. In fact, including disorder amounts to simply adding terms only along the diagonal of the Hamiltonian matrix. To do this we can generate a diagonal matrix with only the disorder terms, which are given by 2.11, acting on each state. Our Hamiltonian them becomes:

\begin{center}
$H = H_{interaction} + H_{disorder}$
\end{center}

And so we have fully generated our Hamiltonian via element-wise construction. We must however make note of the Hilbert space dimensions for the full Hamiltonian: $2^N$. Even for just 16 spins, exact diagonalization calculations for the full Hamiltonian matrix are quite time consuming. However, we can break up our workload and only look at states with total spin set to zero: $S_z = 0$. In this way we only deal with $N$ choose $\frac{N}{2}$ states. This amounts to block diagonalizing the full Hamiltonian and only dealing with the $S_z = 0$ block. To ensure we only work with total $S_z = 0$ states we only choose states with an equal number of 1's and 0's, which gives us our $S_z = 0$ block. While this block is still large, it is significantly less so than the full Hamiltonian, and takes much less time to exact diagonalize. In this way, and if we choose, we could analyze the full Hamiltonian, one block at a time.\\
\bigskip

All our code is written in Python, and we can further reduce our computation time by remembering that we can easily generate sparse Hamiltonian matrices, and use Scipy's sparse eigensolver to only solve for M states closest to energy density 0.5; M is less than the full number of states for the $S_z = 0$ block. In this way we are only dealing with M states to calculate quantities such as entanglement entropy, and the adjacent gap ratio.
\bigskip

\section{Entanglement Entropy Calculation}
\bigskip

Entanglement Entropy is found by determining the eigenvalues of the reduced density matrix, which is found by taking the partial trace over the full density matrix for a given state:\\

\begin{center}
$ \rho_A = Tr_B(\rho)$
\end{center}

Where $\rho_A$ is the reduced density matrix for system A, and $Tr_B(\rho)$ is the partial trace. The entanglement entropy, $S_{AB}$, is found by:\\

\begin{center}
$$ S_{AB} = -\sum_{\lambda_A} \lambda_Aln(\lambda_A)$$ 
\end{center}

where $\lambda_A$ are the eigenvalues of $\rho_A$. Generating the matrix $\rho_A$ for an arbitrary cut is more involved, and for this study we only looked at biparte entanglement entropy calculations; biparte meaning that the system has been 'cut' into two contiguous sections and are denoted as A and B. The construction of $\rho_A$ is the primary concern when finding $S_{AB}$ as the diagonalization $\rho_A$ is rather straightforward given its dimensions. The dimensions for $\rho_{A}$ are simply $2^{N_A}$ where $N_A$ is size of subsystem A, and since $S_{AB} = S_{BA}$ one can always calculate EE from the smaller subsystem, with the maximum matrix size being $2^{\frac{N}{2}}$. The reduced density matrix for a biparte system is easily constructed through use of single value decomposition. To generate $\rho_A$ for each state, we simply reshape the corresponding Hamitonian eigenvector into a $2^{N_A}$ x $2^{N_B}$ matrix, denoted as $M$, and perform the matrix product with $M^{\dagger}$:\\

\begin{center}
$\rho_A = MM^{\dagger}$
\end{center}

where $N_A(N_B)$ is the number of spins in system A(B). However since we only generate block Hamiltonians the eigenstates, of which are not the same dimension as the full Hilbert space, must be expanded to the full Hilbert space before being reshaped. Once expanded, the full states are then used to find $\rho_A$, the eigenvalues of which are used to find the entanglement entropy, $S_{AB}$.

\bigskip
\section{Adjacent Gap Ratio and its Probability Distribution}
\bigskip

The adjacent gap ratio and its probability distribution are used to look at the statistics of the energy levels. The AGR is given by:\\

\begin{center}
$\delta_n = E_{n+1} - E_n\hspace{2cm}r_n = min(\delta_n,\delta_{n+1})/max(\delta_n,\delta_{n+1})$
\end{center}

where $E_n$ are the eigenvalues sorted in ascending order. For the AGR calculation we average $r_n$ for each set of eigenvalues which gives $<r_n>$. We then average $<r_n>$ over the number of $\phi$ and $\phi '$ iterations for a given disorder amplitude which gives, $[<r_n>]$; we will refer to $[<r_n>]$ as $\textbf{r}$, which is the quantity that is plotted. The value of $\textbf{r}$ indicates whether the system is in the ergodic phase given by the GOE limit, $\textbf{r} = 0.568$, or possible MBL phase given by the Poisson limit, $\textbf{r} = 0.386$.\\
\bigskip

We also look at the probability distributions for $\textbf{r}$, P($r$), at different disorder values to show the transition from a GOE distribution to a Poisson distribution for the MBL phase. As we will see, however, for our model the AGR does not obey this behavior.

\newpage

\chapter{Results}

\bigskip
For our model, we explored different system sizes and looked at similar quantities for each one namely, entanglement entropy (EE) and the adjacent gap ratio (\textbf{r}). We also only looked at the states closest to $\epsilon = 0.5$, $\epsilon$ is the energy density, in the total $S_z = 0$ block for our Hamiltonian. We used opened boundary conditions along both $x$ and $y$ directions for all systems.

\section{Entanglement Entropy}

We look at the biparte \textbf{EE} for different values of N, the subsystems of which are denoted as A, and B. The \textbf{EE} calculations are averaged over 30 states, for all N, for 1000 realizations for both N=12 systems, 200 for N=15, and 100 for both N=16 and N=18 systems. The entanglement entropy between subsystems A and B for different values of N are shown in figure 3.1. We see that our model gives low entanglement for higher disorder amplitudes as is characteristic expected for an MBL phase [4], and has been shown in other numerical studies for 2D systems[6].

\begin{figure}[!h]
\begin{center}
\includegraphics[scale=0.4]{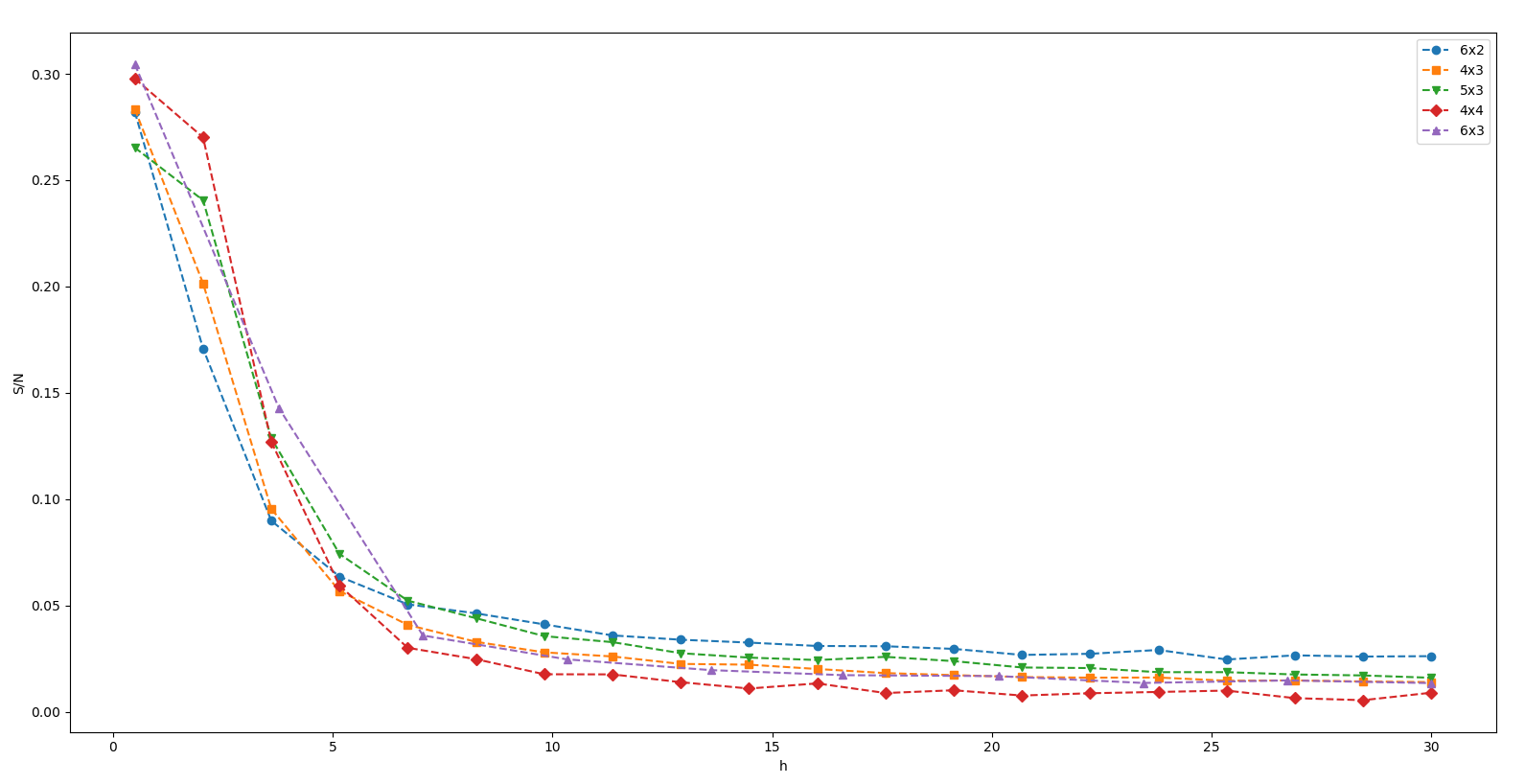}
\caption{\textit{Here we show the entanglement entropy per site for different system sizes, for the quasi-Heisenberg model. Here we see a drop in EE as the system becomes more localized with increasing h, indicating an MBL phase.}}
\end{center}
\end{figure}

If we look at the probability distributions for the EE at different disorder values, as shown in figure 3.2, most of the entropy values approach zero, with a small peak around $ln(2)$, as the system transitions to an MBL phase. The $ln(2)$ peak is expected as this represents some level of entanglement between systems A and B at the boundary of the subsystems.
\bigskip

\begin{figure}[!h]
\begin{center}
\includegraphics[scale=0.3]{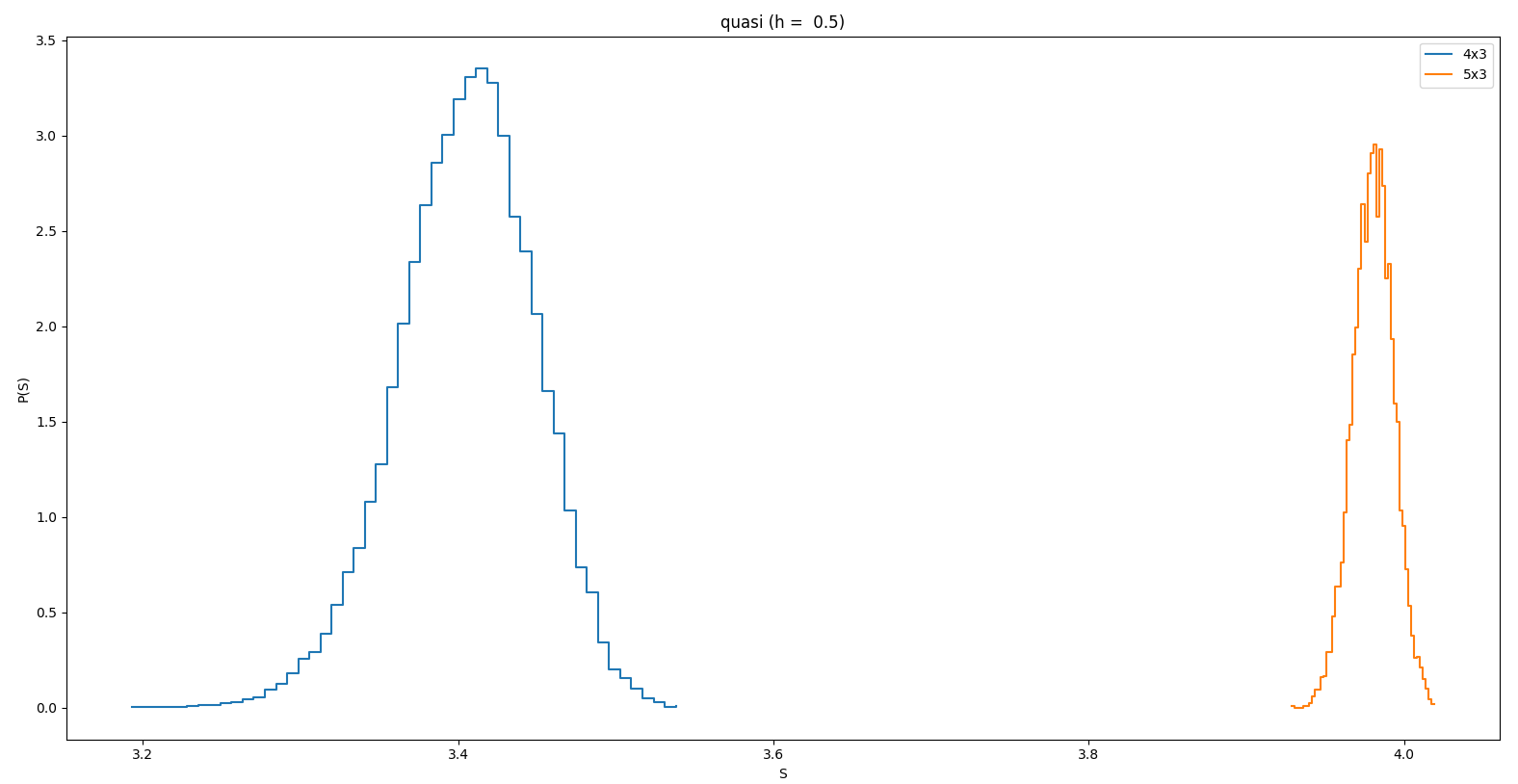}\\
\includegraphics[scale=0.3]{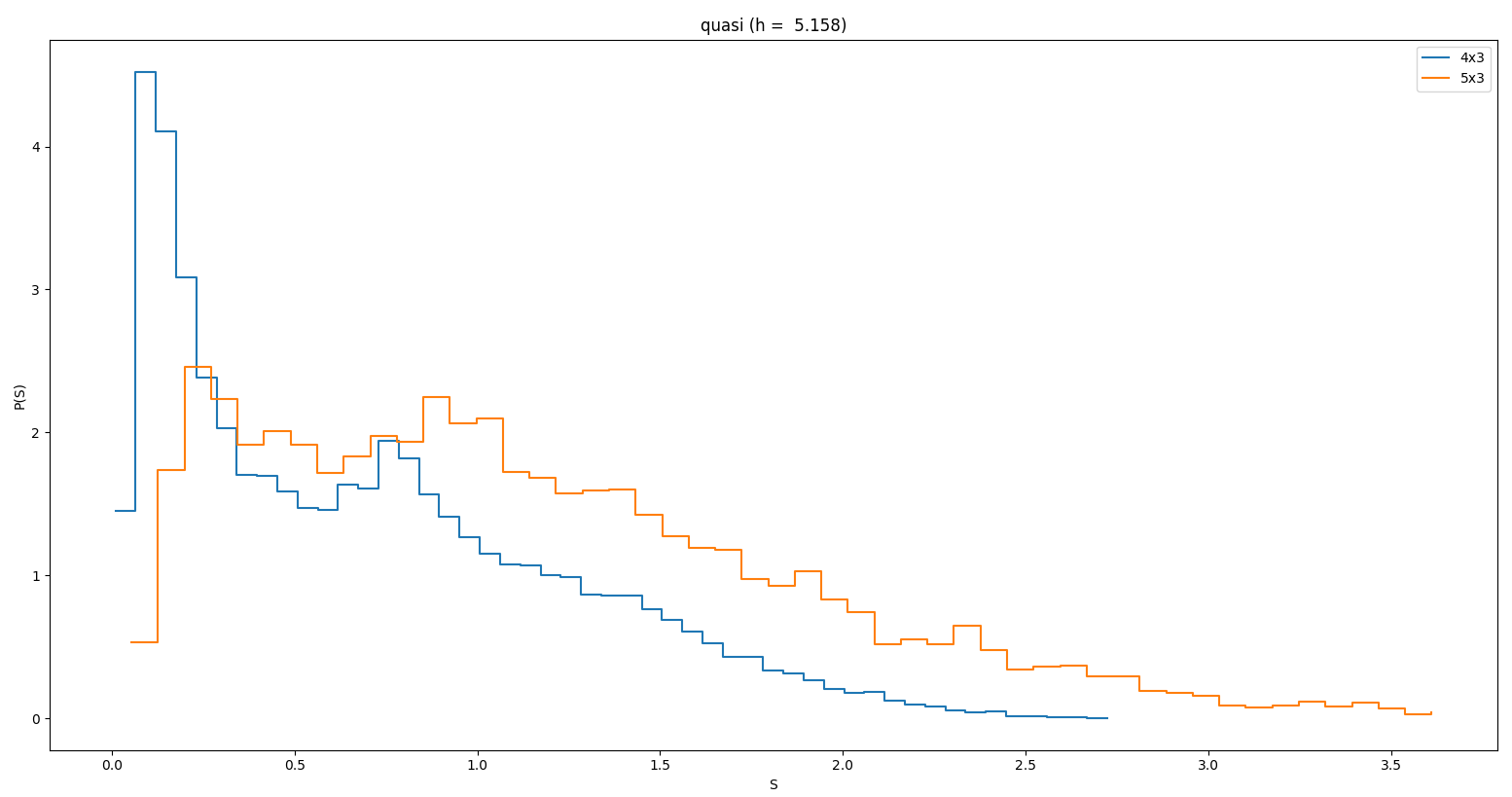}
\includegraphics[scale=0.3]{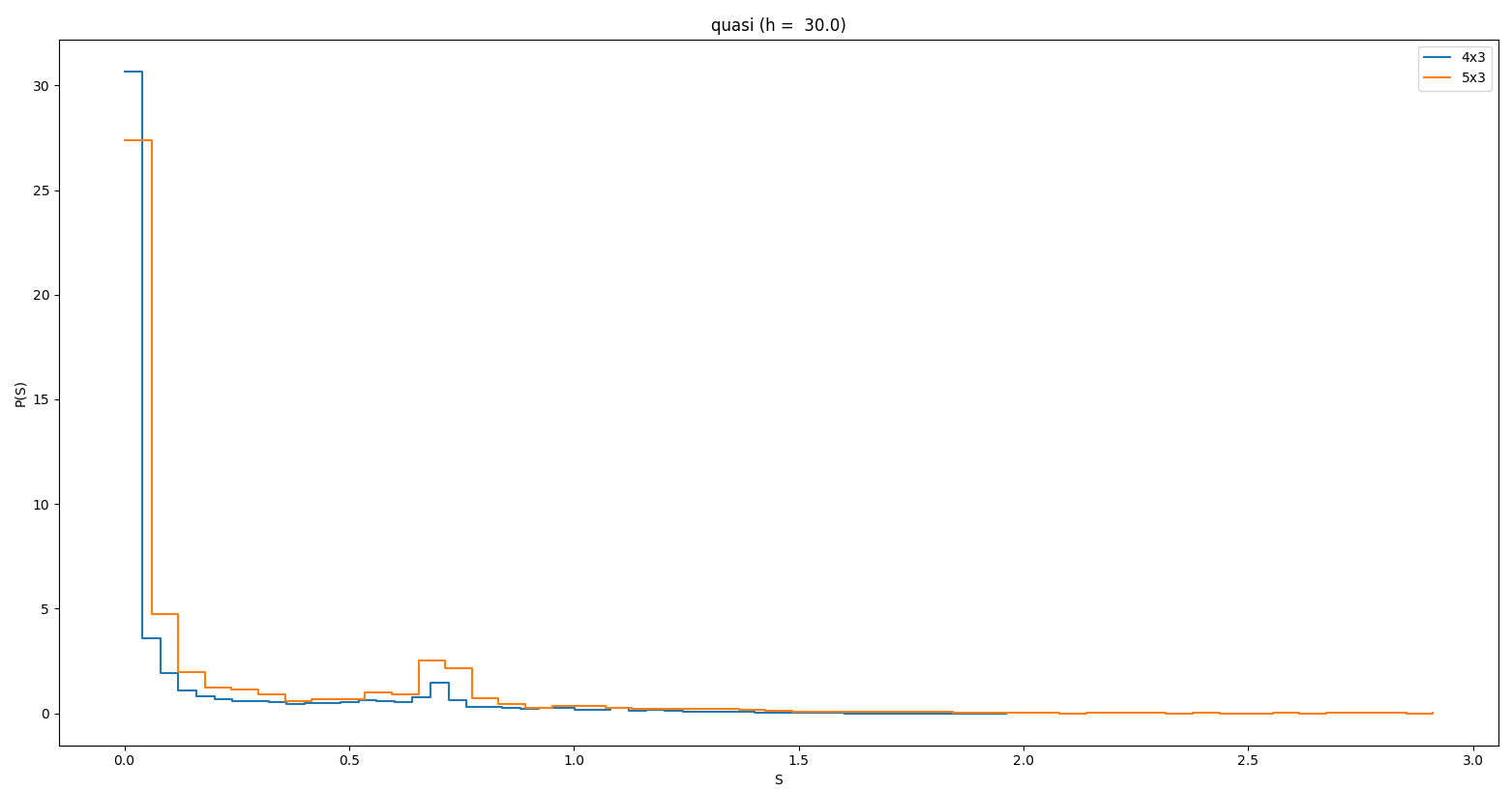}
\caption{\textit{Plot of EE distributions with peaks forming around $S=0$, and $S=ln(2)$ when a system enters the MBL phase, as disorder amplitude is increased. The different plots are for different system sizes, both showing a peak forming around $ln(2)$ for increasing disorder amplitude}}
\end{center}
\end{figure}

The MBL phase is characterized by a reduction in entanglement entropy between subsystems, as seen in this case.

\section{Adjacent Gap Ratio}

As the different systems enter the MBL phase there is an expected transition of adjacent gap ratio values (\textbf{r}), from the GOE limit, 0.5295, to the Poisson limit, 0.386[3]. In fig. 3.3 we calculated a spectral average for the AGR for each disorder amplitude (h), for both quasirandom, and uniform random disorder. For 30 states near energy density 0.5, we calculated \textbf{r} with 1000 realizations for the 4x3 lattice, 200 realizations for the 5x3 lattice, and 100 realizations for the 6x3 lattice.\\

\begin{figure}[h]
\begin{center}
\includegraphics[scale=.3]{AGR_compare_uniform}
\includegraphics[scale=.3]{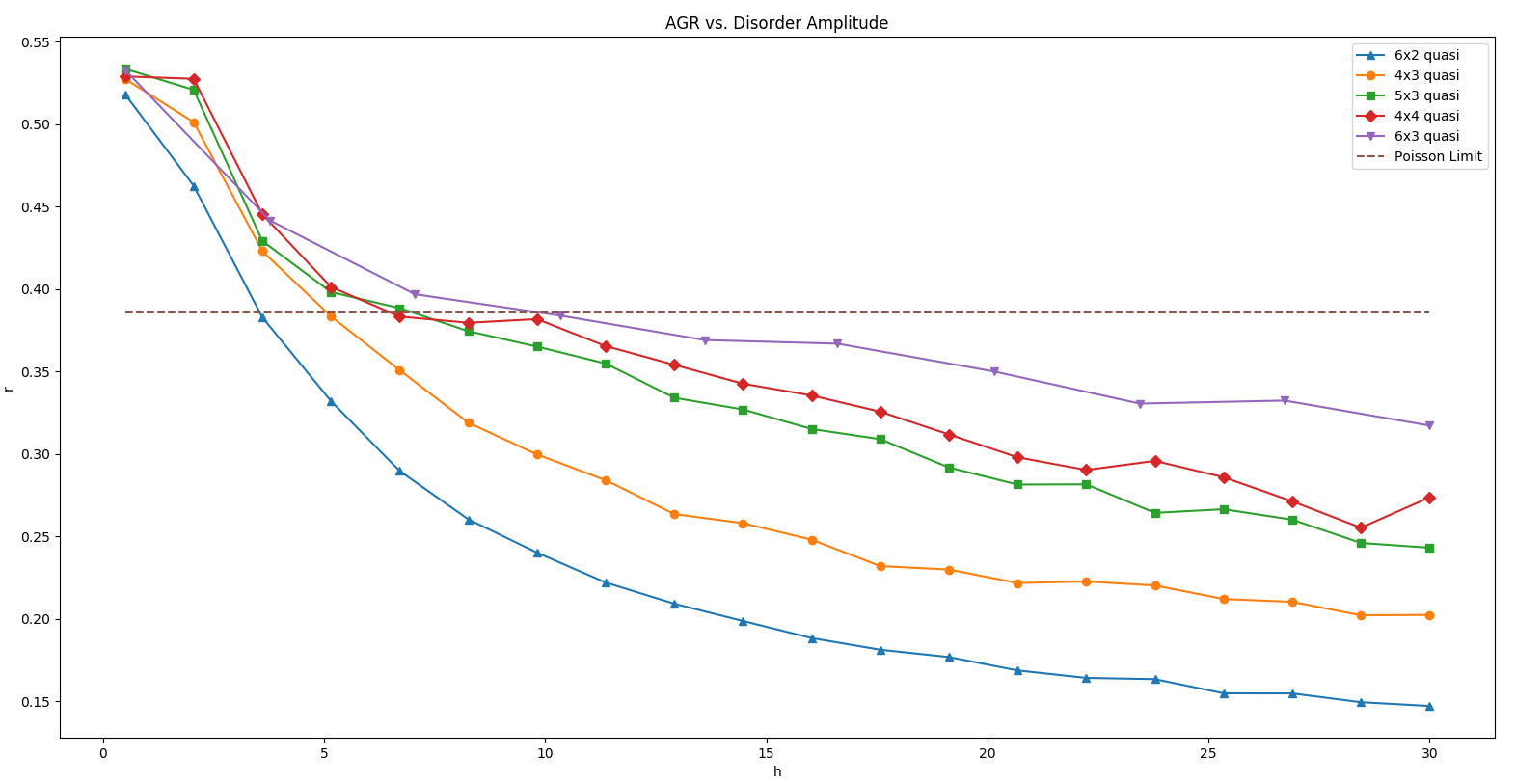}

\end{center}
\caption{\textit{Plots comparing the adjacent gap ratios for the {\normalfont 4x3, 5x3, 6x3} lattices as a function of disorder amplitude; the horizontal line in both is the Poisson limit. The top plot is for quasirandom disorder, and the bottom plot is for uniform disorder. Note: \textbf{r} for the {\normalfont 6x3} system was only calculated for 10 disorder values, and only for quasirandom disorder.}}
\end{figure}

In other 2D systems with quasi random potential terms which exhibit an MBL phase as the disorder amplitude is increased, it has been shown that \textbf{r} approaches the Poisson limit, corresponding to uncorrelated energy levels[5]. Where as for our model, the limit for \textbf{r} extends well below the Poisson limit. Smaller values of \textbf{r} are indicative of an increased number of energy level clustering, with gaps between clusters; that is, neighboring energy levels are close in energy, hence clustering.\\

\begin{figure}
\begin{flushleft}
\includegraphics[scale=.3]{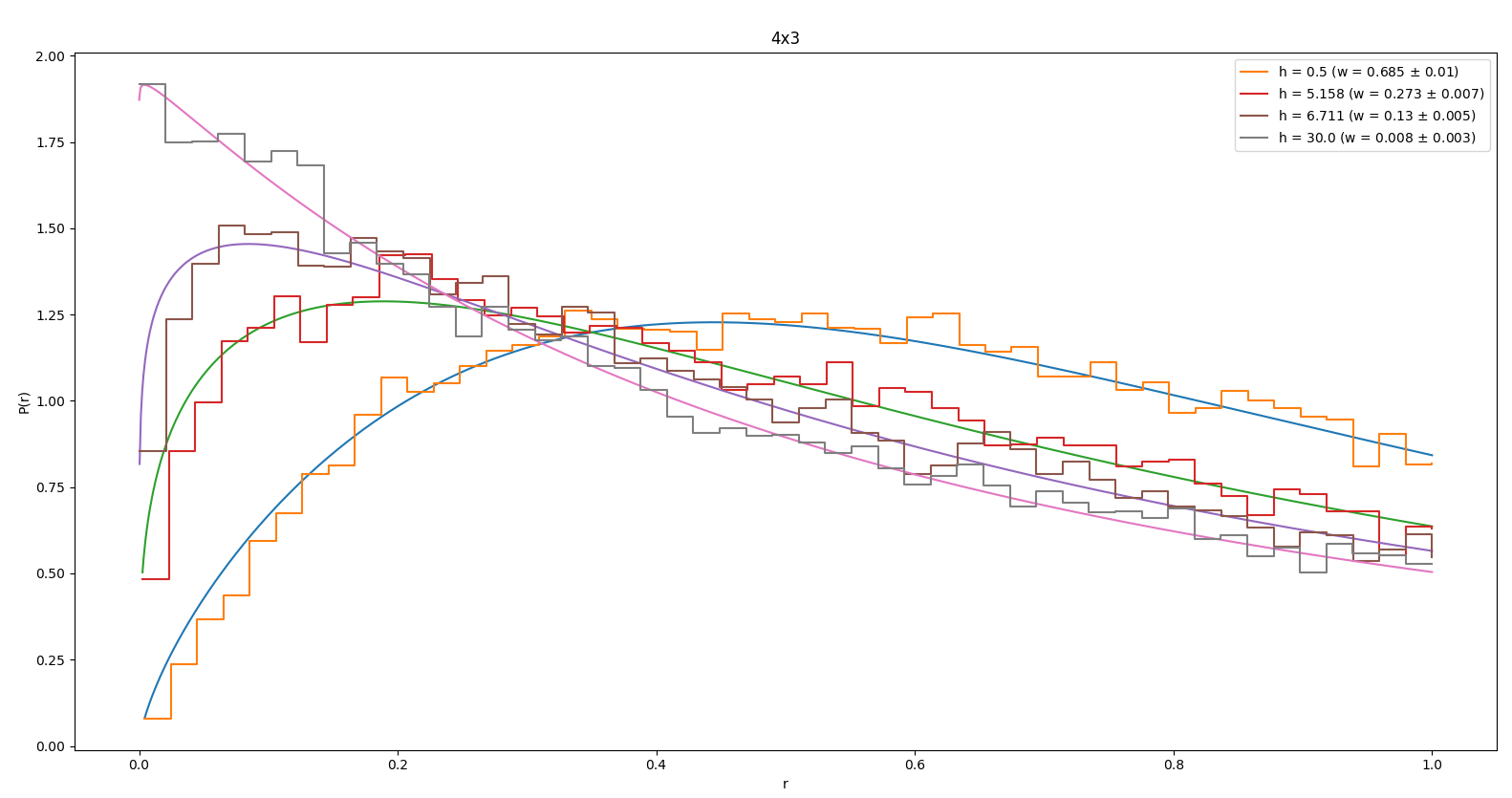}\includegraphics[scale=.3]{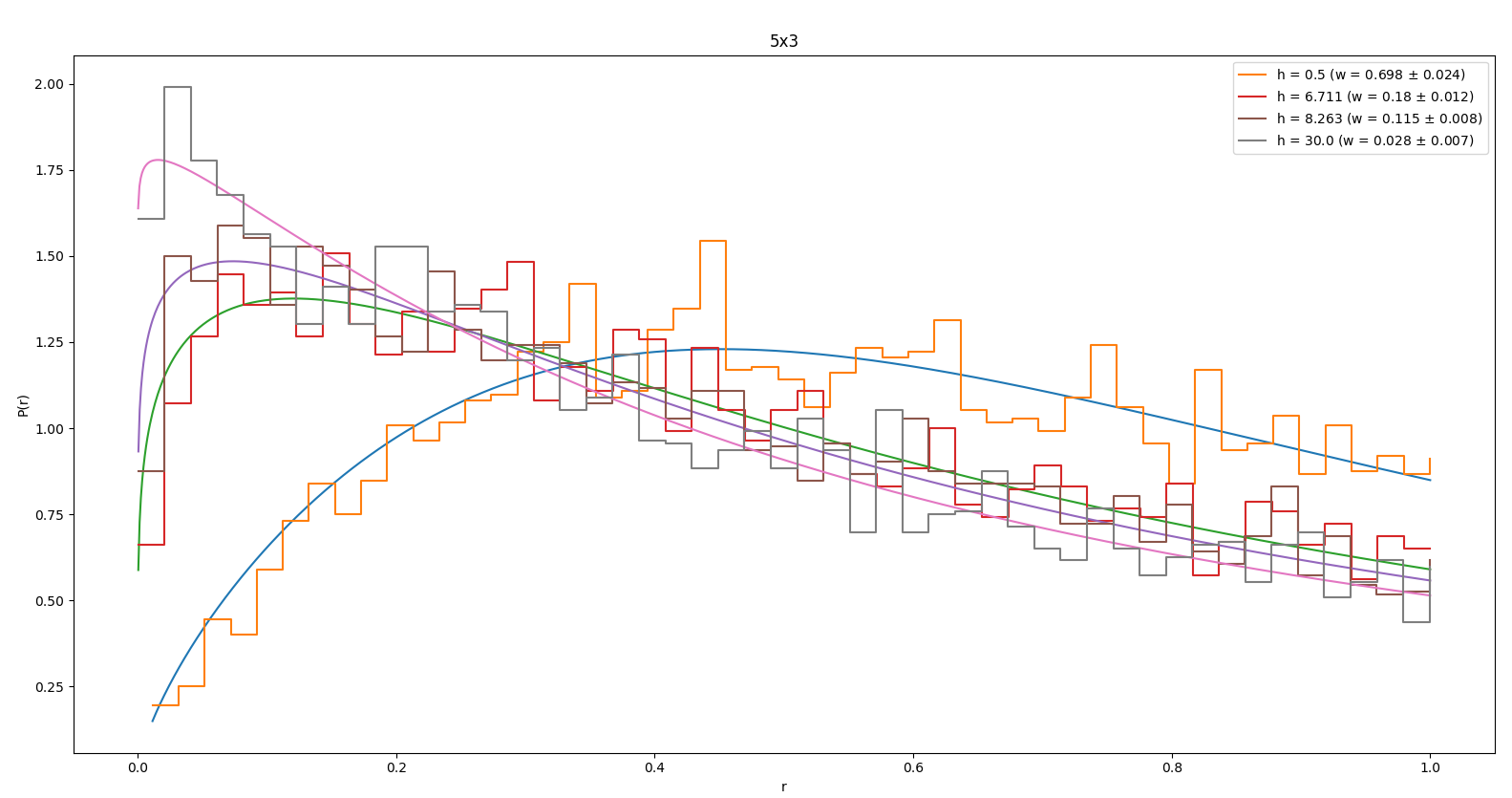}
\end{flushleft}
\caption{\textit{P(r) for different system sizes, with uniform disorder, showing the expected transition from a GOE, to Poisson distribution as we increase the disorder amplitude. We also fit these distributions with eq. 3.2, where $\omega = 1(0)$ gives the GOE(Poisson) distribution.}}

\end{figure}

Whether or not the values of \textbf{r} continue to decrease or approach some limiting value remains unclear. If we look at the probability distributions for \textbf{r}, there is increased energy level clustering as h increases which, is denoted by a shift towards small values for \textbf{r} in plots for P(r).\\

\begin{figure}[!h]
\begin{center}
\includegraphics[scale=.3]{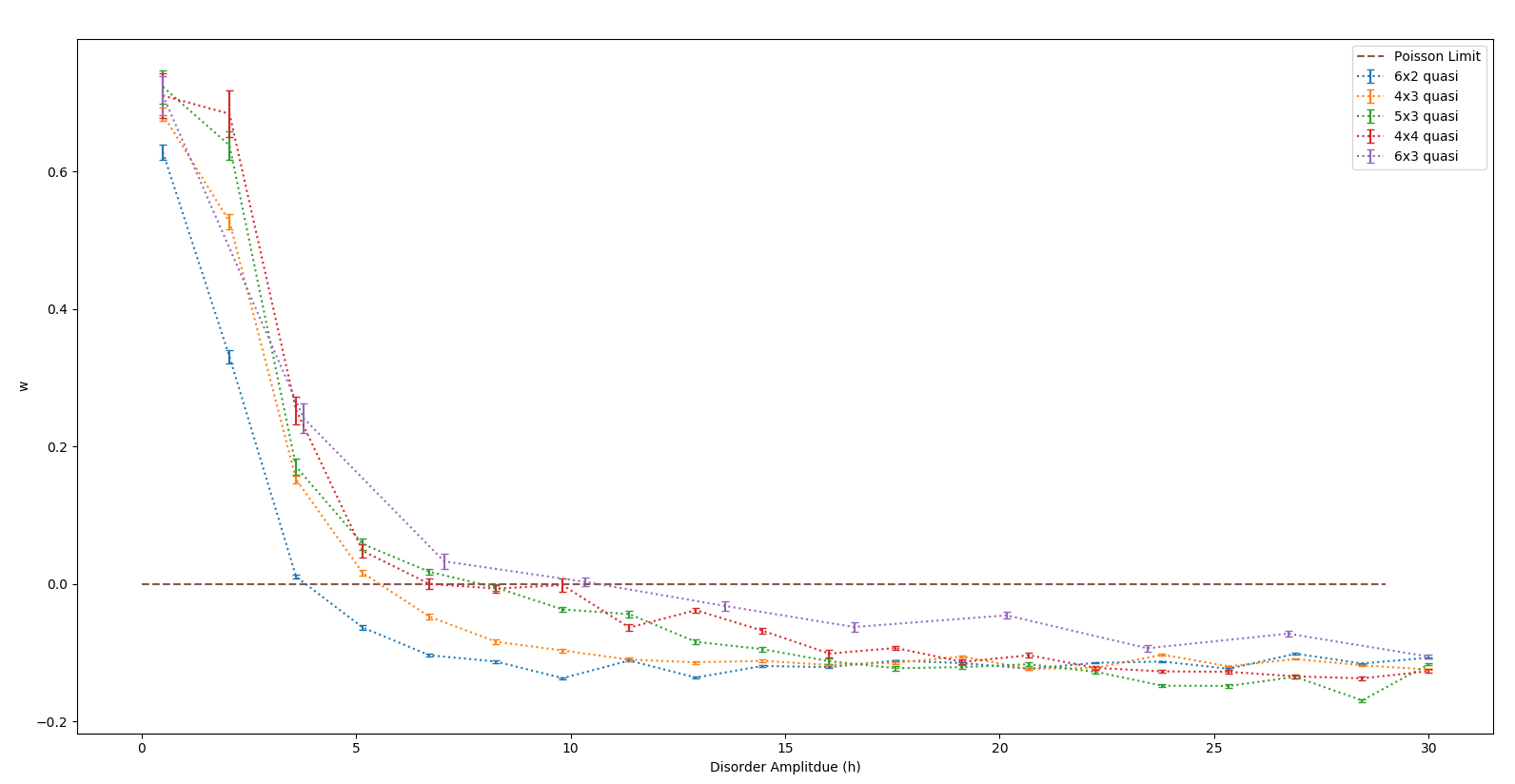}\includegraphics[scale=.25]{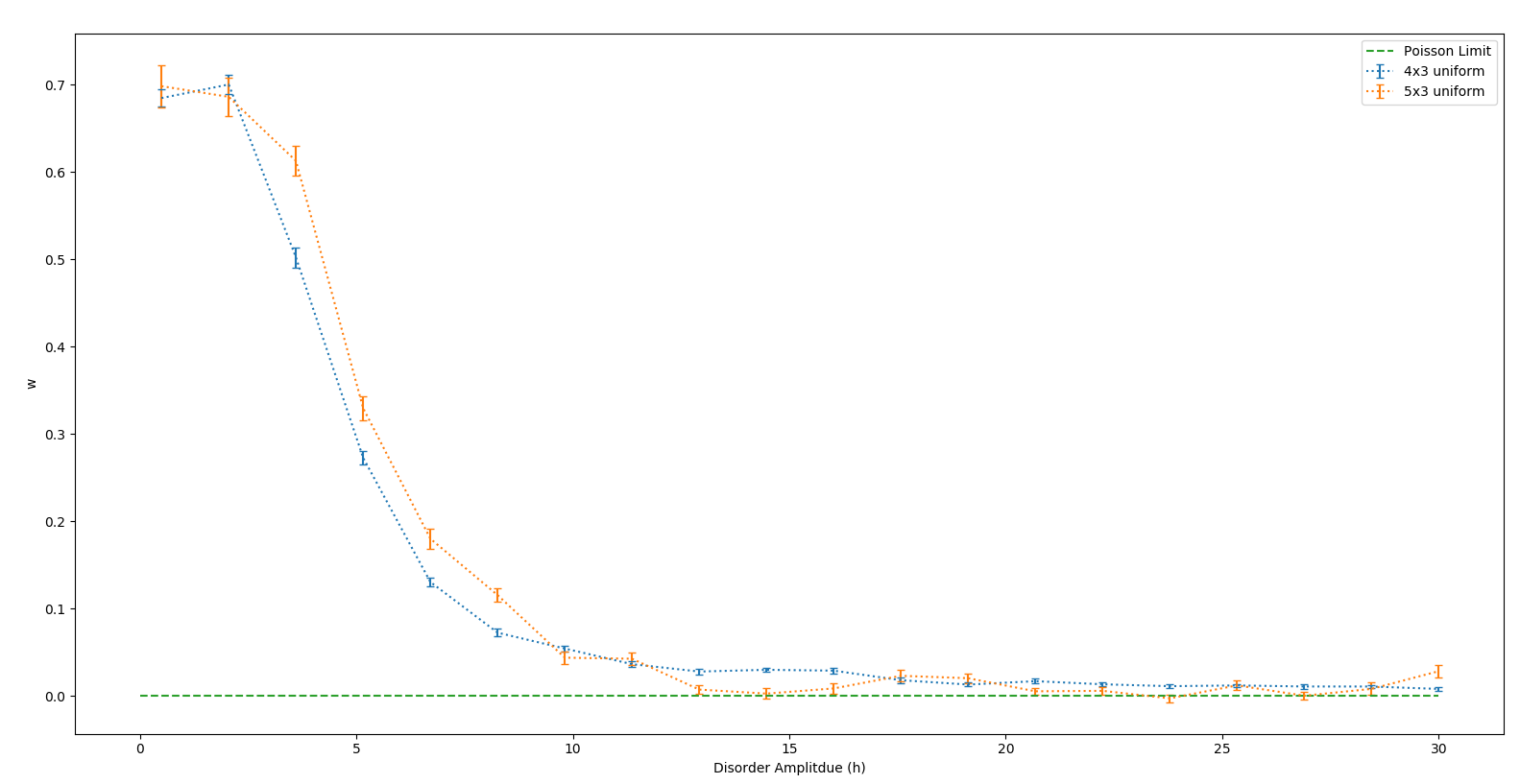}
\end{center}
\caption{\textit{Plot of the fitted Brody parameters as a function of disorder amplitude. The different curves are for different system sizes. The plot on the left is for quasirandom disorder, where the one on the right is for uniform disorder. We see $\omega$ reflect the expected behavior for \textbf{r} in the right plot, and it tends towards its Poisson value ($\omega = 0$).}}
\end{figure}

There are various distributions used to fit P(\textbf{r}) as shown in [7] and [5], but the limits on those only account for \textbf{r} falling between the GOE, and Poisson limits. We need a distribution that can account for some degree of energy level clustering. For this we use the distribution used by [9]: the Brody distribution [8] for energy level spacing P(s,$\omega$), where $\omega$ is the Brody parameter: 

\begin{center}
$P(s,\omega) = A(\omega)s^\omega e^{-\alpha(\omega) s^{\omega +1}}\hspace{0.5cm} (3.1)$
\end{center}

where $A(\omega) = (\omega + 1)\alpha(\omega)$, and $\alpha(\omega) = [\Gamma (\frac{\omega+2}{\omega + 1})]^{\omega + 1}$. The spacing parameter $s$ is given by $\frac{S}{D}$, where $S$ is the difference in energy between levels and $D$ is the average spacing for the energy spectrum. The distribution is well defined for $\omega >-1$, and interpolates between the Wigner surmise, $P(s) = \frac{\pi}{2}e^{-\frac{\pi}{4}s^2}$ with $\omega = 1$, and the Poisson distribution, $P(s) = e^{-s}$ with $\omega = 0$ for energy level spacing [10]. However, a negative value for $\omega$ is indicative of level clustering, for which the distribution is still valid [8][9]. We use the Brody distribution to derive P(r). In following reference [7] we start by deriving P(r') where: 

\begin{center}
$r_n' = \frac{s_n}{s_{n+1}}$, with $s_n = e_{n+1}- e_{n}$\\
\bigskip

$r_n = \frac{min(s_n,s_{n-1})}{max(s_n,s_{n-1})} = min(r_n',\frac{1}{r_n'})$\\
\bigskip

$P(r') == \int P(s_1,s_2)\delta(r' - \frac{s_1}{s_2})ds_1ds_2 \rightarrow P(r') = \int_{0}^{\infty} P(r's_2,s_2)s_2ds_2$

\end{center}

where $r_n$ is the standard adjacent gap ratio introduced by Oganesyan and Huse [3], and $P(s_1,s_2)=\rho(e_1,e_2,e_3)$ is the probability density for three consecutive energy levels, $e_1 \leq e_2 \leq e_3$. Note that $P(r) = 2P(r')\Theta(1-r')$, where $\Theta(1-r')$ is the Heavyside step function, and restricts the distribution to $r' = [0,1]$. The remainder of the derivation is in Appendix I, but distribution for P(r) is found to be:

\begin{center}
$P(r) = \frac{2(\omega +1)r^{\omega}}{(1+r^{\omega + 1})^2} \hspace{0.5cm}(3.2)$
\end{center}   

For $\omega = 0$ the energy level spacing is Poisson, and indeed we recover the same P(r) as [3] for the Poisson distribution: $P(r) = \frac{2}{(1+r)^2}$. We fit our data for P(r) to 3.2 using the least squares method, and we plot $\omega$ as a function of disorder amplitude, $h$. 

\begin{figure}[!htb]

\includegraphics[scale=.3]{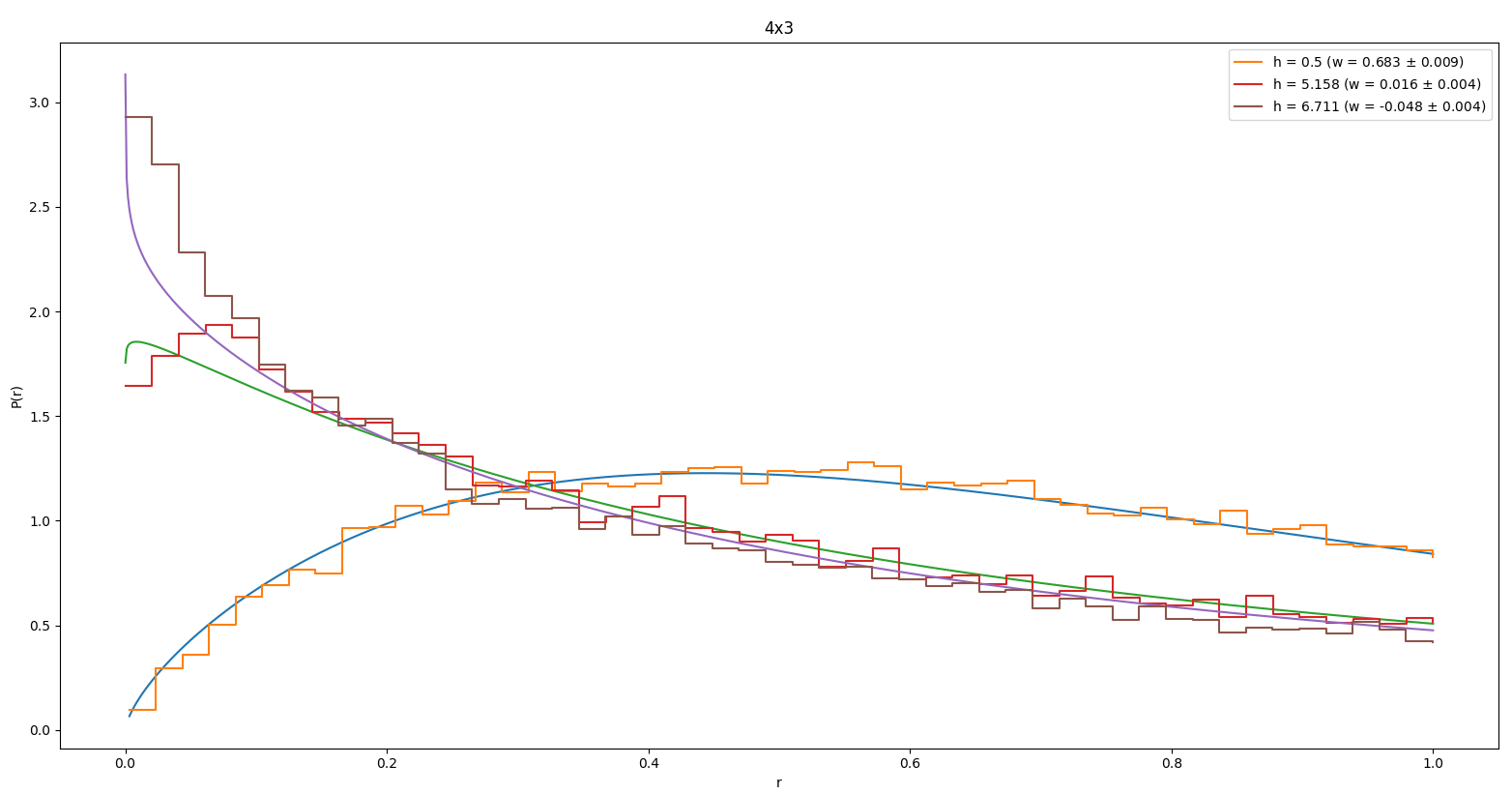}
\includegraphics[scale=.25]{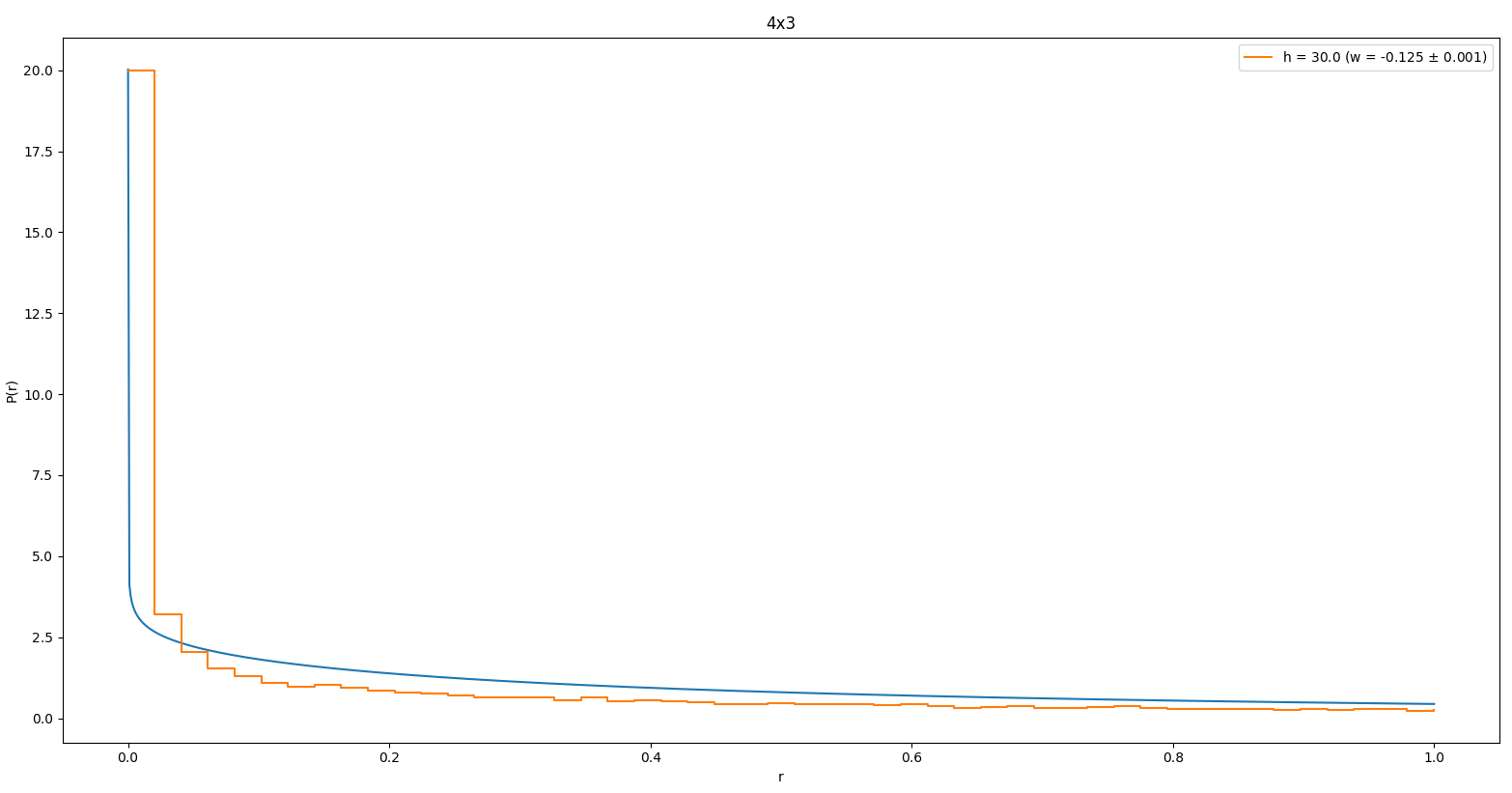}

\includegraphics[scale=0.3]{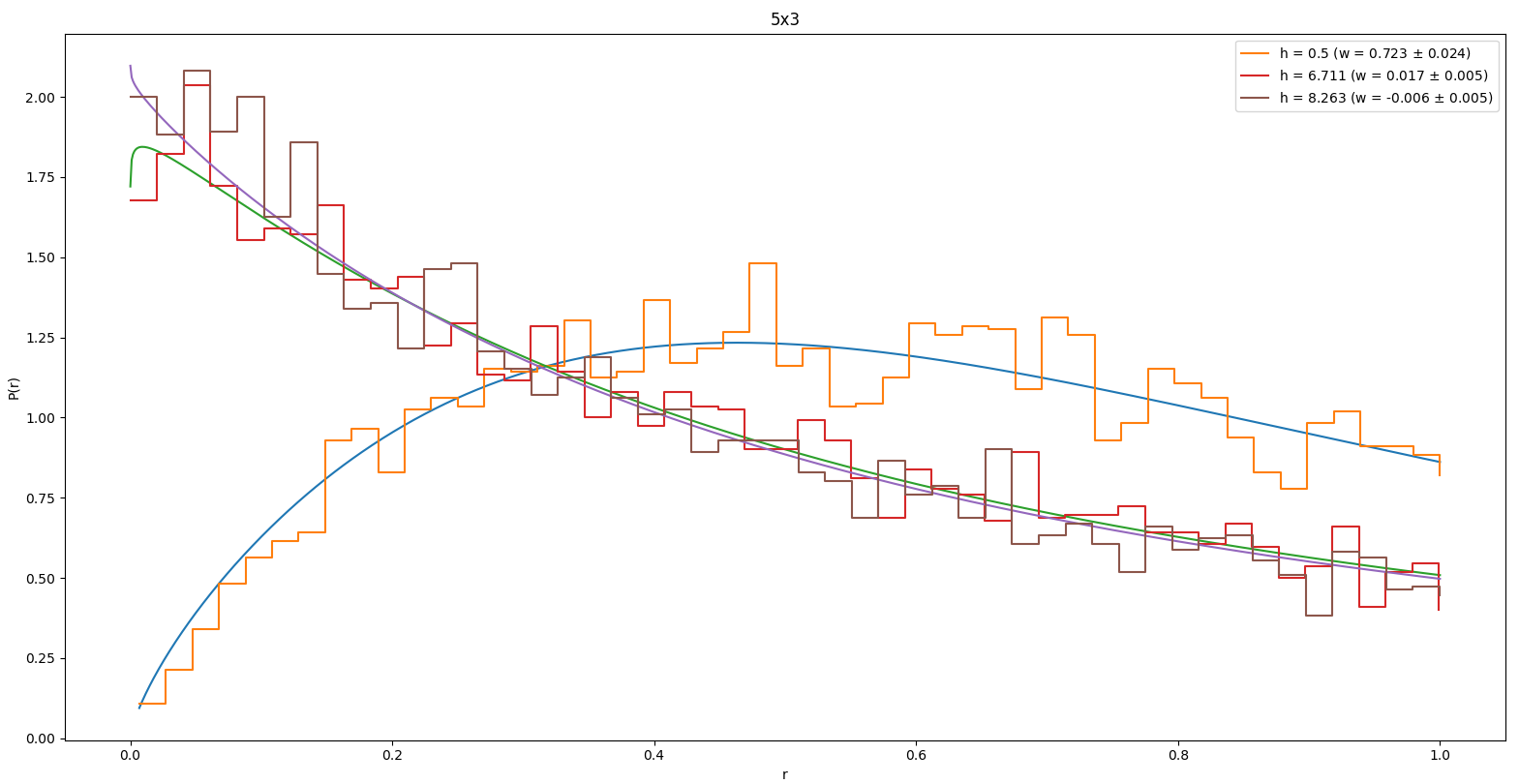} 
\includegraphics[scale=.25]{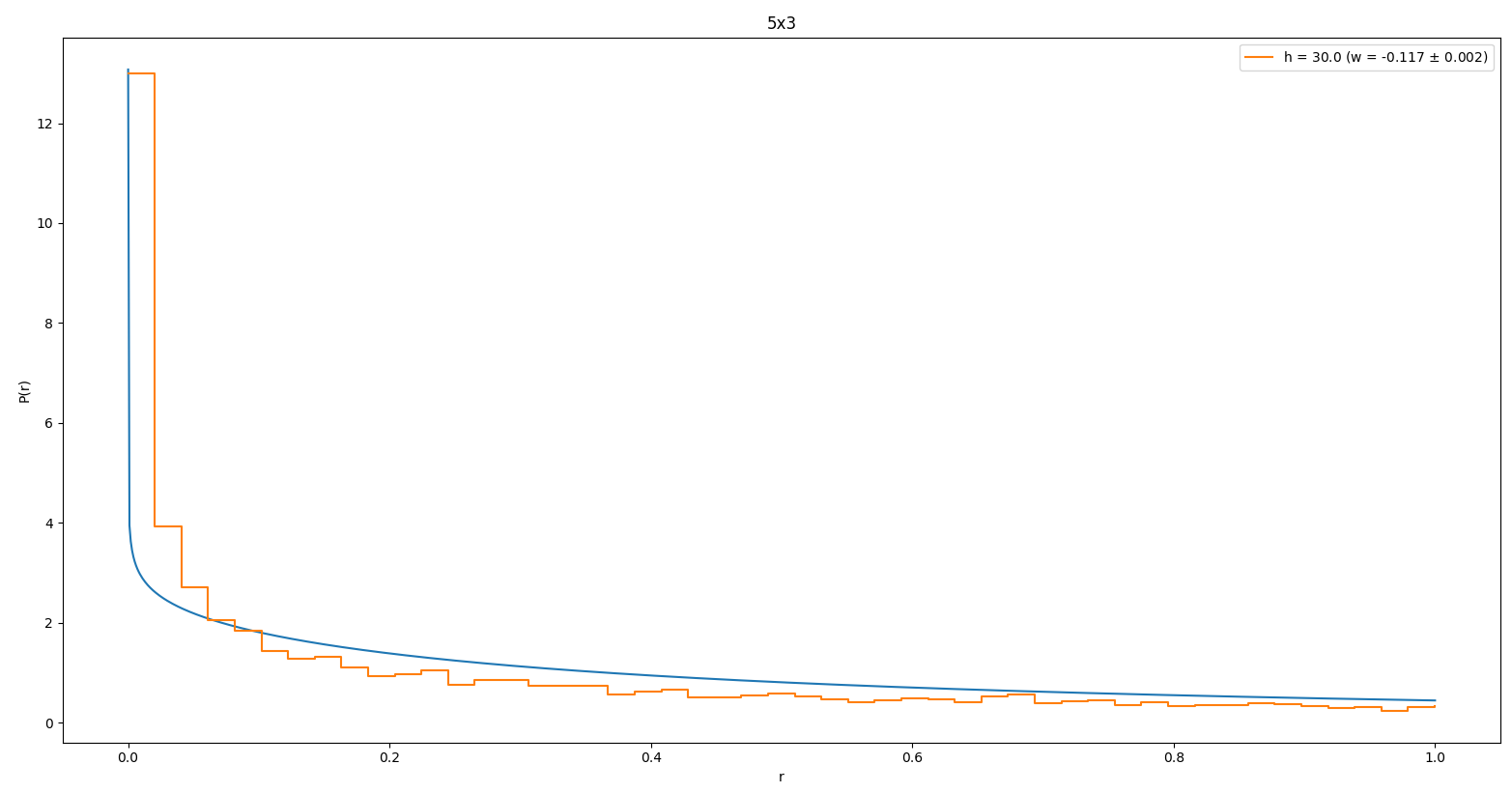}

\includegraphics[scale=.3]{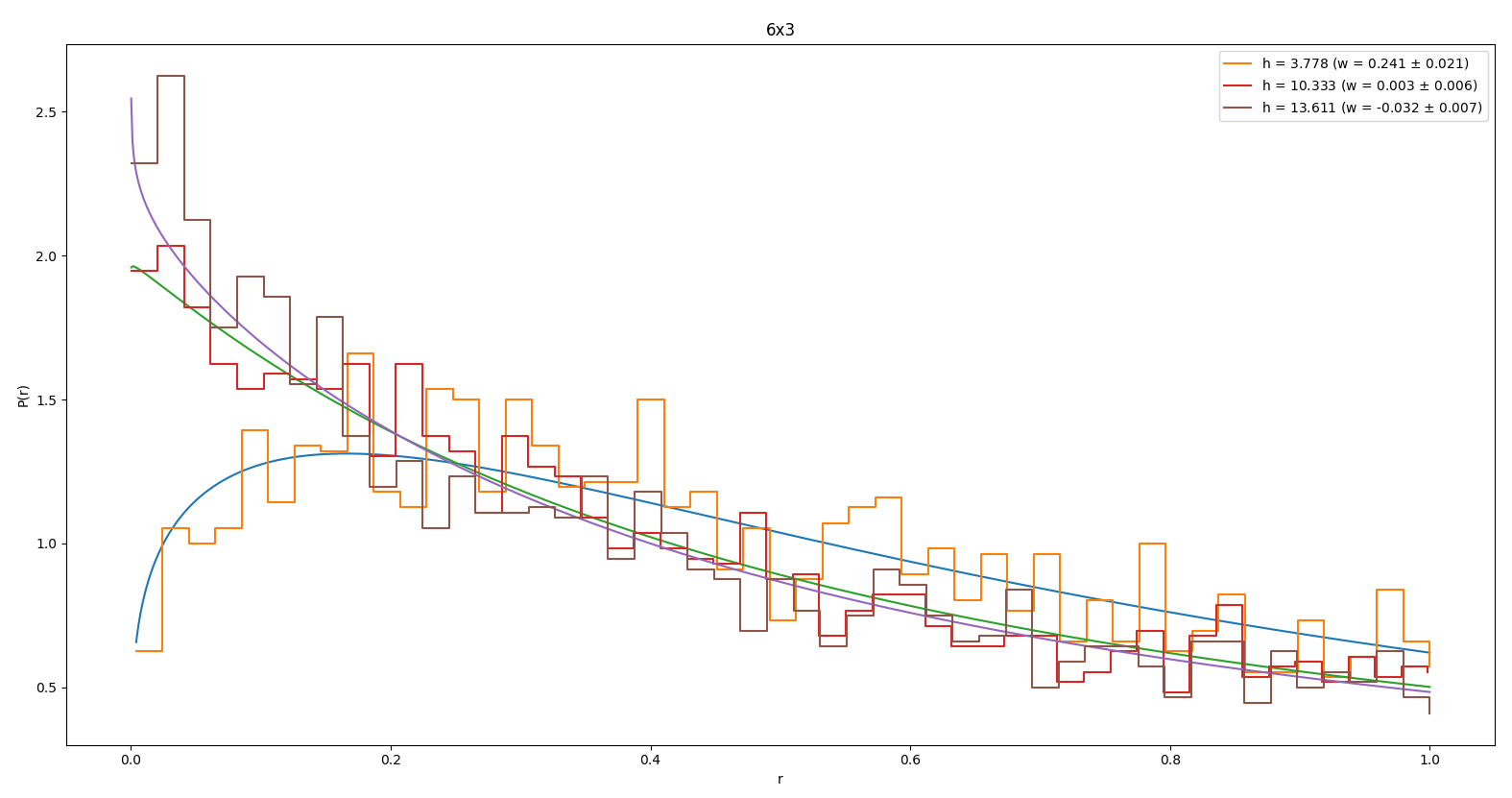}
\includegraphics[scale=.25]{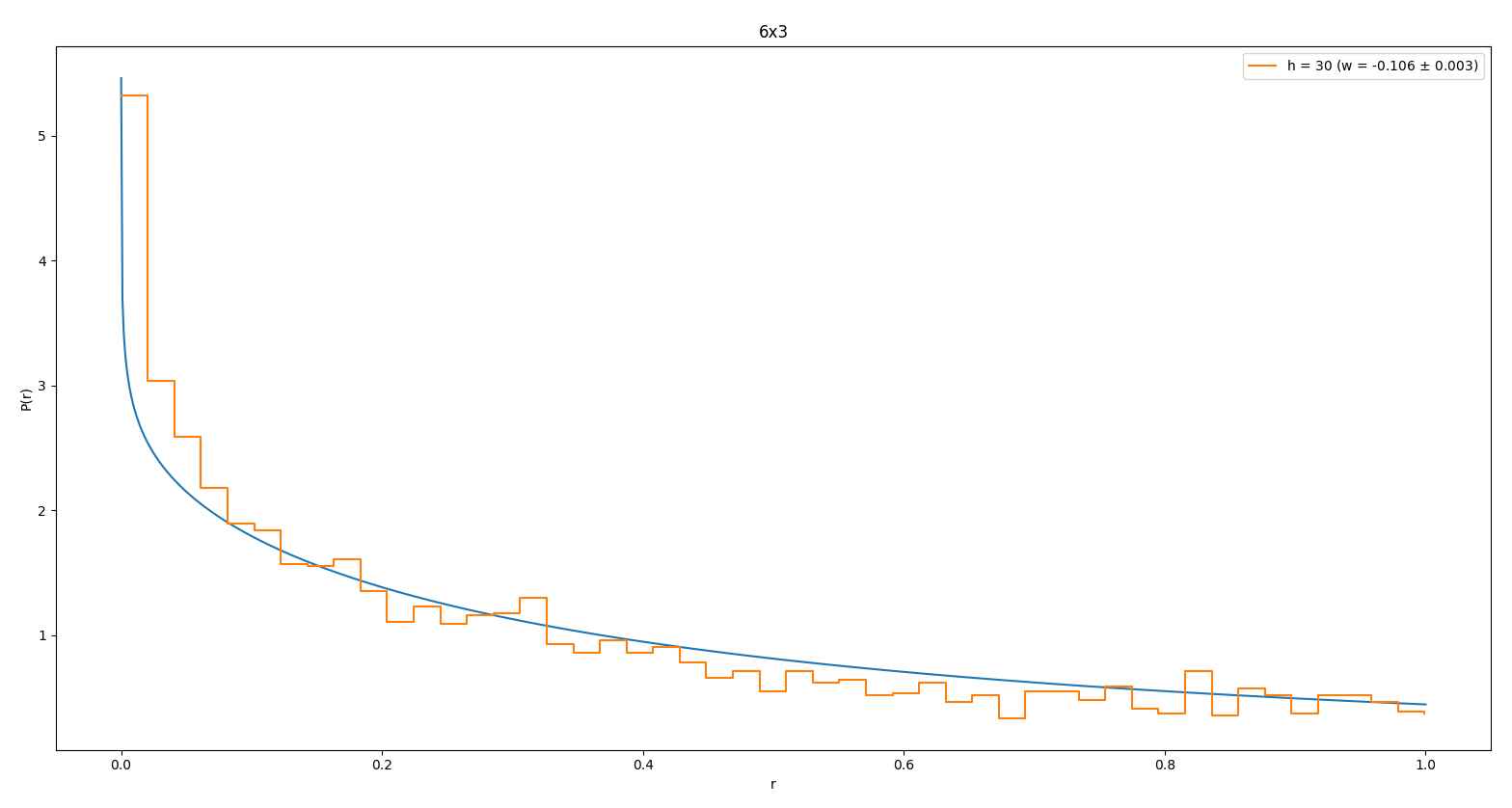}
\caption{\textit{P(r) for different nx3 lattice sizes, with each plot showing P(r)for different disorder amplitudes with corresponding Brody parameter ($\omega$). The right column shows P(r) for the extreme disorder value h=30.0. For the left column, we show a transition from a GOE , to Poisson, to a level clustering distribution. We notice in the extreme values for h, the distribution sharply peaks for small values of r, indicating most levels are clustered for higher disorder values. We also note that as total system size increases, the clustering at h=30.0 is less extreme.}}
\end{figure}

\begin{figure}[!htb]
\includegraphics[scale=.3]{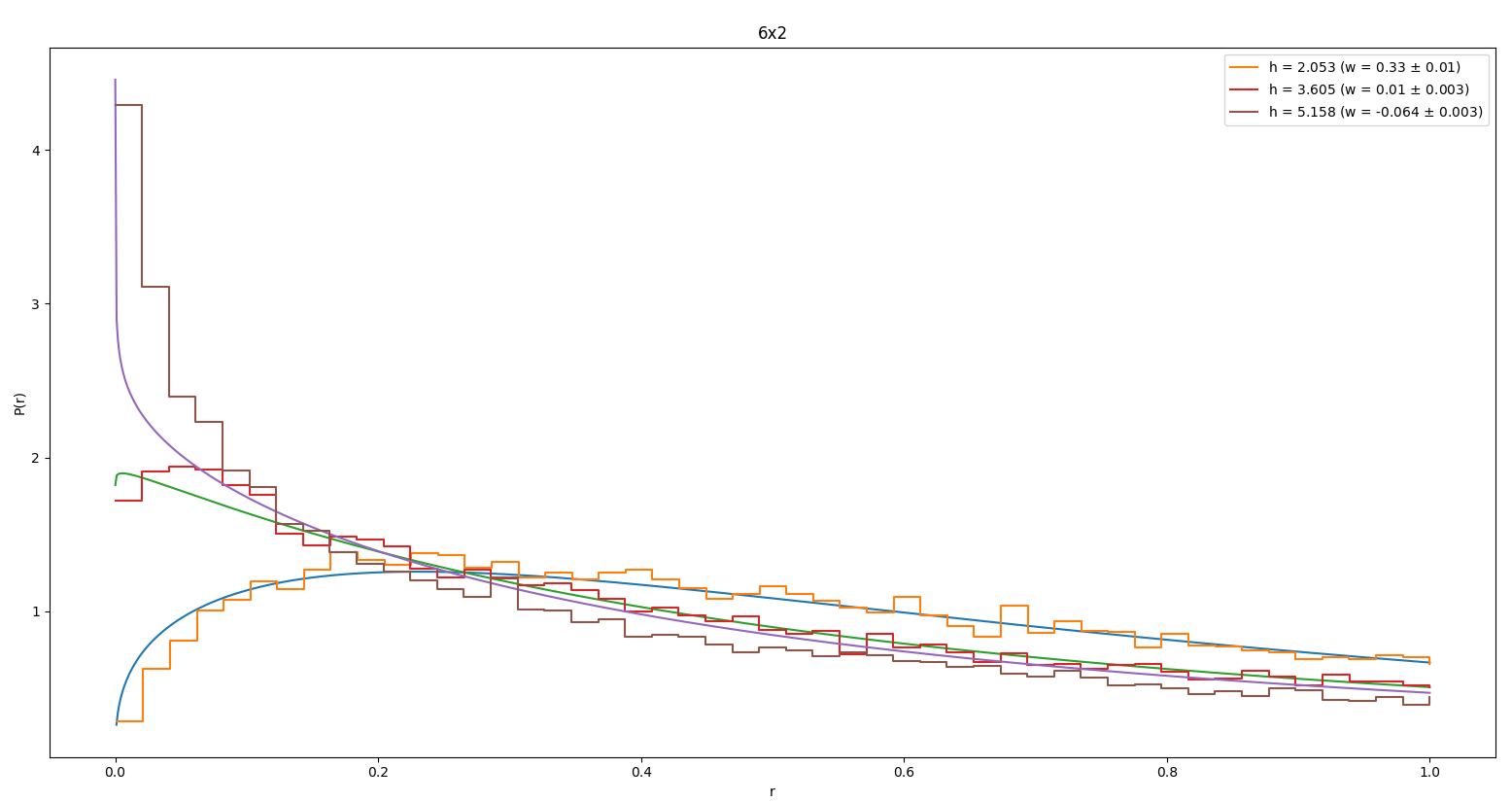}
\includegraphics[scale=.25]{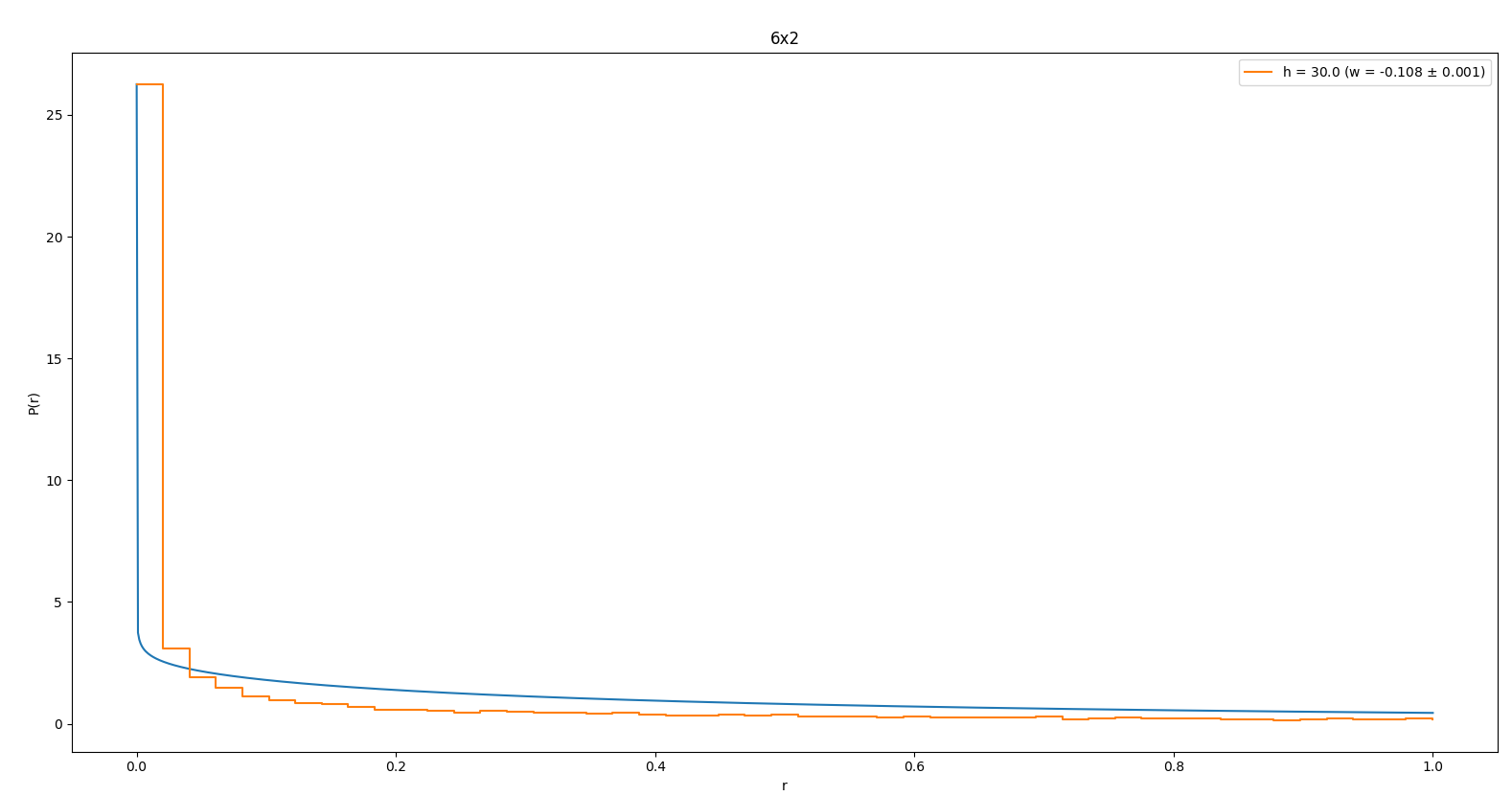}

\includegraphics[scale=.3]{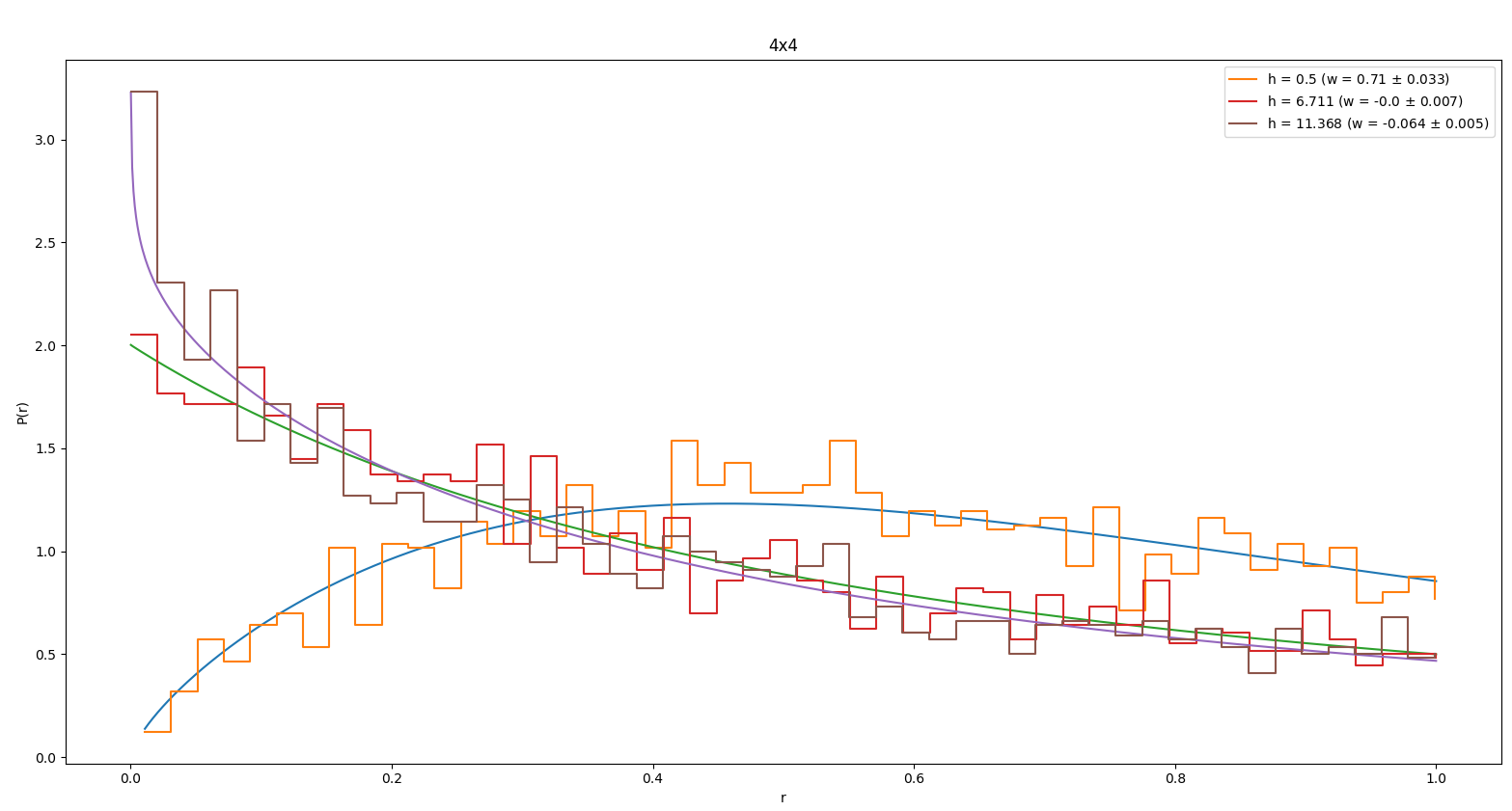}
\includegraphics[scale=.25]{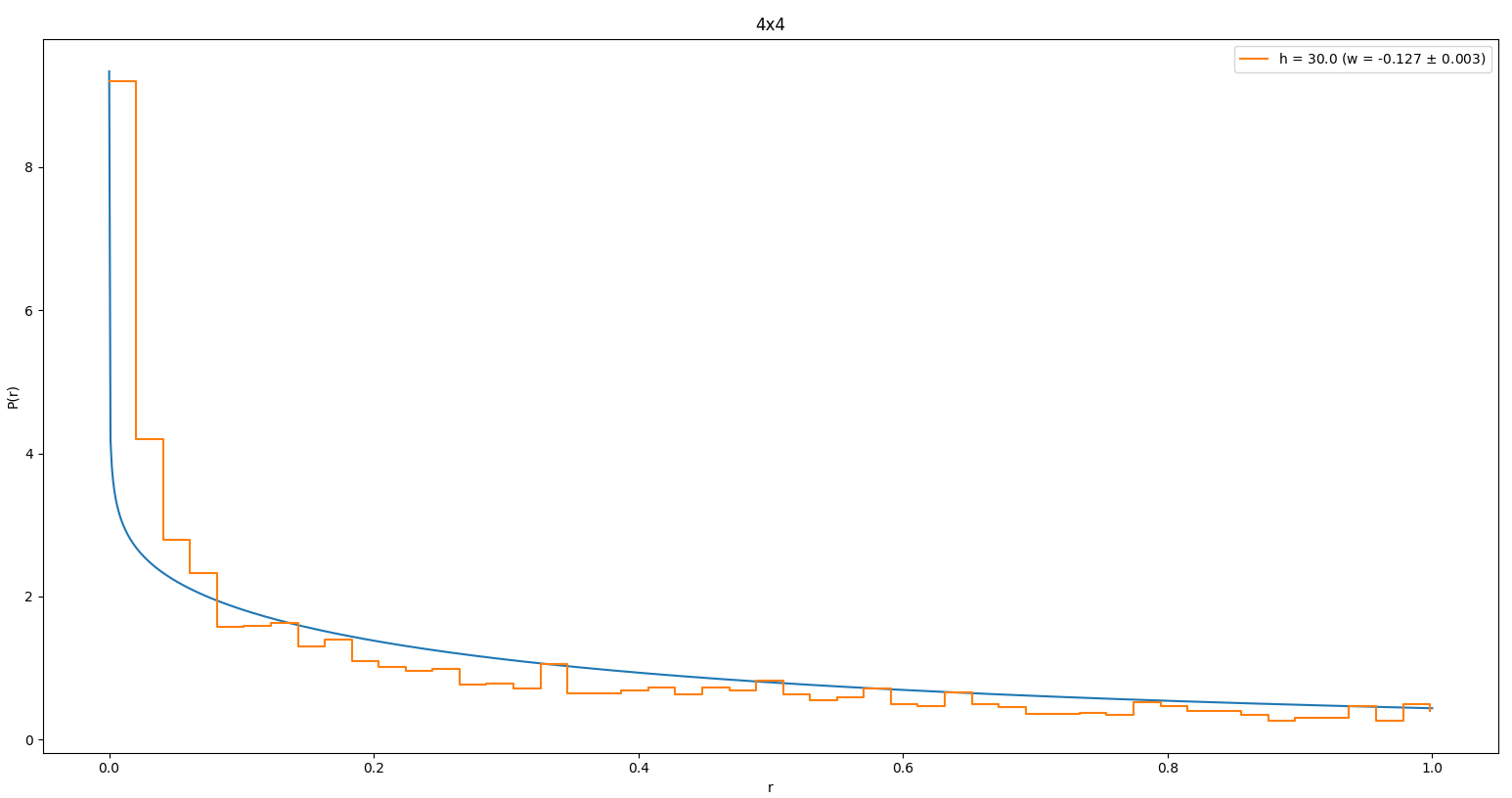}
\caption{\textit{P(r) for 6x2, and 4x4 lattice, with each plot showing P(r) for different disorder amplitudes with corresponding Brody parameter ($\omega$). The right column shows P(r) for the extreme disorder value h=30.0. We still see a transition from a GOE, to Poisson, to level clustering distribution as h increases, and again we the clustering is less extreme for h=30.0, for larger system sizes. Despite a different geometry to that of the nx3 lattice, the behavior is similar, indicating an independence of geometry.}}
\end{figure}

Next, we fit eq. 3.2 to the distributions for when $\omega$ changes from positive to negative as well as extreme values for $h$; each $\omega$ has a corresponding $h$ given as $h ,\omega \pm \sigma_{\omega}$: shown in figures 3.6, and 3.7.\\
\bigskip

For each system size, the transition for when \textbf{r} falls below the Poisson limit, 0.386, is qualitatively reflected in the shape of the corresponding P(r), and quantitatively reflected in the sign of the Brody parameter, $\omega$. That is, as \textbf{r} dips below the Poisson limit, the sign of $\omega$ shifts from positive to negative. The $h$ value for this transition($h_0$) varies based on system size, increasing for larger systems. For $N= 12$, $ 5.158 \leq h_0 \leq 6.711$, for $N=15$, $6.711 \leq h_0 \leq 8.263$, and for $N=18$, $ 10.333 \leq h_0 \leq 13.611$. Also, the transition away from the Poisson distribution ($\omega < 0$) is not as rapid for the larger system sizes. In looking at the extreme plot for P(r), we see that energy level clustering increases dramatically as h is taken to its extreme value for different system sizes.

\section{Discussion}

Previous numerical results for the 2D disordered-Heisenberg model[6], and in 1D quasirandom-Heisenberg model[12] have both shown a decrease in entanglement entropy as the system transitions to an MBL phase. The entanglement entropy results for our 2D model are in agreement with the characteristics for an MBL phase to occur [1][4].\\

The reduced value for \textbf{r} as the disorder amplitude is increased indicates level clustering, and the probability distributions for the AGR also reflect the unexpected values of \textbf{r}. As the disorder amplitude is increased there is a corresponding increase in the amount of 'small' values for individual $r_n$, and thus leading to an overall lowering of \textbf{r} below the Poisson limit. It has been theorized that since the energy levels are uncorrelated, \textbf{r} should approach the Poisson value in the localization phase as $N \rightarrow \infty$[3]. As we increase the system size, we do see that \textbf{r} tends back toward the Poisson limit in Fig. 1.3, which could indicate finite size effects. However, exact diagonalization limits our system size to no more than $18$, and so we cannot explicitly show if \textbf{r} does indeed converge back to its theorized Poisson value with increased system size.\\

However, similar models ([5][6]) have shown that larger system sizes are not needed to demonstrate a Poisson distribution for \textbf{r}, indicating something unique in our model. Further inspection of the disorder terms: 

\begin{center}
$$H_{disorder} = \sum_i S_i^z[cos(2\pi\sqrt{2}n_x + \phi ') + cos(2\pi\sqrt{2} n_y + \phi )]$$
\end{center}

reveals that the disorder field can be completely separated into two separate fields along $x$ and $y$ directions. Therefore there are a certain number of spins which share the same $n_x$($n_y$) value, e.g. each spin in a certain row will share the same disorder value given by, $h_n^x = hcos(2\pi\sqrt{2}n_x + \phi )$.

\begin{center}
$(h_1^x,h_1^y)\bullet $\hspace{1cm}$\bullet (h_1^x,h_2^y)$\\

\vspace{1cm}
$(h_2^x,h_1^y)\bullet $\hspace{1cm}$\bullet (h_2^x,h_2^y)$
\end{center} 

We can quickly see that each spin's $h_n^x(h_n^y)$ term has a corresponding term with another spin. We can rewrite the disorder Hamiltonian as:

\begin{center}
$$H_{disorder} = \sum_n \sum_m S_{mn}^ih_n^x + \sum_m \sum_n S_{mn}^ih_m^y$$
\end{center}

where n is a sum over rows, and m is a sum over columns. The degeneracies arise when comparing different spin configurations, for when both summations for different configurations match we see a degenerate pair of states. Only there's no restriction in only having pairs, and so multiple configurations can have the same energy. To illustrate this we use a 2x3 lattice and compare two separate spin configurations:
\bigskip

Configuration 1:

\begin{center}
$(h_1^x,h_1^y)\uparrow $\hspace{1cm}$\downarrow (h_1^x,h_2^y)$\hspace{1cm}$\downarrow(h_1^x,h_3^y)$\\

\vspace{1cm}
$(h_2^x,h_1^y)\downarrow $\hspace{1cm}$\uparrow (h_2^x,h_2^y)$\hspace{1cm}$\uparrow(h_2^x,h_3^y)$
\end{center} 
\bigskip

Configuration 2:

\begin{center}
$(h_1^x,h_1^y)\downarrow $\hspace{1cm}$\uparrow (h_1^x,h_2^y)$\hspace{1cm}$\downarrow(h_1^x,h_3^y)$\\

\vspace{1cm}
$(h_2^x,h_1^y)\uparrow $\hspace{1cm}$\downarrow (h_2^x,h_2^y)$\hspace{1cm}$\uparrow(h_2^x,h_3^y)$
\end{center}
\bigskip

To compare the two, we only need to keep track of the sign for each spin; we assign a value of 1 to spin up and -1 to spin down, and add up all values along the same direction for each row(column).
\bigskip

\begin{minipage}{0.5\textwidth}
	\begin{center}
		Configuration 1\\
		\begin{tabular}{ c c c|c }
		0 & 0 & 0 \\
		\hline
		1 & -1 & -1 & -1 \\
		-1 & 1 & 1 & 1 \\
		\end{tabular}
	\end{center}
\end{minipage}%
\begin{minipage}{0.5\textwidth}
	\begin{center}
		Configuration 2\\
		\medskip
		
		\begin{tabular}{ c c c|c }
		0 & 0 & 0\\
		\hline
		-1 & 1 & -1 & -1\\
		1 & -1 & 1 & 1\\
		\end{tabular}
	\end{center}
\end{minipage}
\bigskip

We see that both spin configurations have total $S_z = 0$, and while the configurations are different, they have the same energy. Each spin configuration represents some basis state, e.g. the above configurations for our methods would represent the basis states: $\ket{100011}$, and $\ket{010101}$ with the first three spins representing the top row, and the remaining three representing the bottom. We should note that there is no indication of dimensionality from the basis states themselves, and what matters is that they are two distinct spin configurations. This, however, is just one instance of many degenerate states, and among the $2^N$ basis states there are several degenerate states. If we restrict ourselves to only total $S_z = 0$ basis states, we only need to compare 20 states for $N=6$, which is what was done for the above example. However, analytically determining which basis states are degenerate remains an open problem, as the above process must be done iteratively and quickly becomes intractable for humans.\\

The level clustering we observe comes from the degeneracies in the eigen-energies from the disorder terms, and for large disorder amplitudes, $\frac{J}{h}<< 1$, the eigen-energies for our model are dominated by those from the disorder Hamiltonian. The degeneracy is lifted from the Heisenberg interaction perturbing the eigen-energies of the disorder Hamiltonian, and we see clustering around the aforementioned degenerate energies. However as $J \simeq h$, the eigen-energies of the system are no longer dominated by the disorder Hamiltonian, and so we expect a decrease in level clustering. We would therefore expect \textbf{r} to transition from the GOE limit, to the Poisson limit, and then reduce further as $h$ is  continually increased, to when the energy levels begin to cluster. In using the Brody parameter to represent different distribution, from figure 3.5, we see the transition from a GOE like distribution($\omega \simeq 1$), to a Poisson-like distribution ($\omega \simeq 0$), to one that is indicative of level clustering ($\omega < 0$).\\

In comparing the nx3 lattice, to the 6x2, and 4x4 lattices, it seems that the level clustering behavior, indicated by figures 3.4 and 3.6, depends only on system size, and not so much on different geometries. It also seems based on figure 3.4, that the over all \textit{fraction} of states which exhibit clustering diminishes. As the total number of states increases as $2^N$ we would expect an absolute increase in the number of clustering states. However, \textbf{r} tending back towards its Poisson value with increasing N, coupled with the fact that \textbf{r} is spectral average, indicates that the overall fraction of clustered states diminishes. Preliminary results show that the fraction of degenerative states increases with system size, however  the interaction term also increases with size resulting in an overall reduction of clustered states.\\

We have determined that our model is unique in that we observe level clustering. This clustering is brought about as a consequence of being able to completely separate the disorder term into x and y directions. While the behavior of \textbf{r} for our model is unusual, it could very well be a finite size effect, and our EE still indicates our system becomes localized with increasing $h$. Should \textbf{r} tend back towards its Poisson value for $N>18$, then our model would demonstrate expected characteristics for an MBL phase.\\
\bigskip

One open question is how does this clustering affect the critical disorder value ($h_c$) for which the system transitions from and ergodic to an MBL phase. To analyze this transition we would look at the EE to determine ($h_c$), and compare it to the ($h_c$) for other models. However, we would have to run exact diagonalization calculations again at a higher resolution around the transition point to determine ($h_c$) for our model. It would also be interesting to explore the dynamics of our system, namely spin relaxation times. A similar model to ours has been experimentally realized in [11], and their system had unusually long relaxation times. It would be of great interest to see if our model also shows similarly long relaxation times.

\newpage
\chapter*{References}
\addcontentsline{toc}{chapter}{References}

\bigskip
[1] R. Nandkishore, D.A. Huse (2015), \textit{Many body localization and thermalization in quantum statistical mechanics}, Annual Review of Condensed Matter Physics, Vol. 6: 15-38\\

[2] O. Lychkovskiy (2013), \textit{Dependence of decoherence-assisted classicality on the ways a system is partitioned into subsystems}, Phys. Rev. A 87, 022112\\

[3] V. Oganesyan, D.A. Huse (2007), \textit{Localization of interacting fermions at high temperature}, Phys. Rev. B 75, 155111 (2007)\\

[4] D. A. Abanin, E. Altman, I. Bloch, M. Serbyn (2019),\textit{Ergodicity, Entanglement and Many-Body Localization}, arXiv:1804.11065 [cond-mat.dis-nn]\\

[5] D. Wiater, J. Zakrewski (2018), \textit{Impact of geometry on many-body localization}, Phys. Rev. B 98, 094202\\

[6] E. Baygan, S.P. Lim, D.N. Sheng (2015), \textit{Many-body localization and mobility edge in a disordered Heisenberg spin ladder}, arXiv:1509.06786 [cond-mat.str-el]\\

[7] Y.Y. Atas, E. Bogomolny, O. Giraud, G. Roux (2013), \textit{Distribution of the Ratio of Consecutive Level Spacings in Random Matrix Ensembles}, Phys. Rev. Lett. 110, 084101\\

[8] T.A. Brody (1973), \textit{A Statistical Measure for the Repulsion of Energy Levels}, Lett. Nuovo Cimento (1973) 7: 482. https://doi.org/10.1007/BF02727859\\

[9] J.A. Scaramazza, B.S. Shastry, E.A. Yuzbashyan (2016), \textit{Integrable matrix theory: Level statistics}, Phys. Rev. E 94, 032106\\

[10] M.L. Mehta (1967), \textit{Random Matricies and the Statistical Theory of Energy Levels}, New York, NY, Academic Press\\

[11] P. Bordia, H. L\"{u}schen, S. Scherg, S. Gopalakrishnan, M. Knap, U. Schneider, I. Bloch (2017), \textit{Probing Slow Relaxation and Many-Body Localization in Two-Dimensional Quasiperiodic Systems}, Phys. Rev. X 7, 041047.\\

[12] M. Lee, T.R. Look, D.N. Sheng, S.P. Lim (2017),\textit{Many-Body Localization in Spin Chain systems with Quasiperiodic Fields}, Phys. Rev. B 96, 075146\\

[13] S.Iyer, V. Oganesyan, G. Refael, D.A. Huse (2013),\textit{Many-Body localization in quasiperiodic system}\\

[14] J.J Sakurai, J. Napolitano (1994), \textit{Modern Quantum Mechanics, 2nd Edition}, San Francisco, CA, Addison-Wesley\\
\newpage
\chapter*{Appendix I}
\addcontentsline{toc}{chapter}{Appendix I}

\bigskip
In following reference [7] we continue the derivation for P(r) where:\

\begin{center}
$P(r') = \int_{0}^{\infty} P(r's_2,s_2)s_2ds_2$
\end{center}

Where, $P(s_2)$ is the distribution for energy level spacing for whatever that may be. For the Poisson distribution $P(s_2) = e^{-s_2}$, but in this case $s_2$ is simply a label and we can drop the subscript. Also, $P(r's)$ is simply the same distribution for $P(s)$, but every $s$ is replaced with $r's$. We next note that $P(r's,s) = P(r's)P(s)$; our $P(s)$ is the Brody distribution $P_B(s)$ [8] given as:

\begin{center}
$P_B(s) = A(\omega)s^{\omega}e^{-\alpha(\omega)s^{\omega + 1}}$ where $A = A(\omega) = (\omega + 1)\alpha(\omega)\hspace{0.5cm}\alpha = \alpha(\omega)$
\end{center}

It then follows that:

\begin{center}
$P(r') = \int_{0}^{\infty} P(r's_2,s_2)s_2ds_2  = \int_{0}^{\infty} A^2(r's)^{\omega}e^{-\alpha (r's)^{\omega+1}}s^{\omega}e^{-\alpha s^{\omega+1}}sds$\\
\bigskip

$P(r') = A^2r'^{\omega}\int_{0}^{\infty}s^{\omega}s^{\omega+1}e^{-\alpha s^{\omega+1}(1+r'^{\omega+1})}ds = A^2r'^{\omega}\int_{0}^{\infty}s^{\omega}s^{\omega+1}e^{-\alpha s^{\omega+1}R(\omega)}ds$ 
\end{center}

Where $R(\omega ) = (1+r'^{\omega +1})$. We make the simple u-substitution of $u = s^{\omega +1}, du = (\omega +1)s^\omega ds$:

\begin{center}
$P(r')= \frac{A^2r'^{\omega}}{\alpha^2R(\omega )^2(\omega +1)}\int_{0}^{\infty} ue^{-u}du = \frac{(\omega +1)^2\alpha^2r'^\omega}{\alpha^2R(\omega)^2(\omega +1)}$
\end{center}

After cancellations and noting that $P(r) = 2P(r')\Theta(1-r')$ we have:

\begin{center}
$P(r) = \frac{2(\omega +1)r^\omega}{(1+r^{\omega +1})^2}\hspace{0.5cm}r = [0,1]$
\end{center}

\end{document}